\documentclass[ aps, showpacs, showkeys, nofootinbib, floatfix, superscriptaddress]{revtex4}

\usepackage{amsfonts}

\usepackage{amssymb}
\usepackage{amsmath}
\usepackage{graphicx}
\usepackage{epstopdf}

\begin{document}
\title{Research on high-frequency quasi-periodic oscillations in black bounce-type spacetime}
 \author{Jianbo Lu}
 \email{lvjianbo819@163.com}
 \affiliation{Department of Physics, Liaoning Normal University, Dalian 116029, P. R. China}
 \author{Shining Yang}
 \email{yangshining1996@163.com}
 \affiliation{Department of Physics, Liaoning Normal University, Dalian 116029, P. R. China}
 \author{Yuying Zhang}
 \affiliation{Department of Physics, Liaoning Normal University, Dalian 116029, P. R. China}
 \author{Liu Yang}
 \affiliation{Department of Physics, Liaoning Normal University, Dalian 116029, P. R. China}
\author{Mou Xu}
 \affiliation{Department of Physics, Liaoning Normal University, Dalian 116029, P. R. China}

\begin{abstract}
  This paper investigates the high frequency quasi-periodic oscillations (HFQPOs) phenomenon around the black bounce-type (BBT) spacetime using the resonance models. We calculated the location of the innermost stable circular orbit (ISCO) for different types of celestial bodies, and derived the expression for the epicyclic frequencies of test particles. The results show that the BBT spacetime possesses unique observational characteristics, where the ordering of epicyclic frequencies varies with the regularization parameter $a$, enabling the excitation of low-order resonances and producing stronger observational signals. Using parametric and forced resonance models, we compared theoretical results with the observed 3:2 twin-peak HFQPOs in microquasars (GRO 1655-40, XTE 1550-564, GRS $1915+105$ ), analyzed the formation mechanisms of HFQPOs, constrained the parameters of the BBT model, and explored the possible types of celestial objects corresponding to microquasars. The study indicates that, certain parametric resonance conditions (e.g., $n=1, 2$) lead to traversable wormhole models in BBT that closely align with observations. And forced resonance corresponding to BH or wormhole models can be verified through observations. These results deviate from the data fits of the original black-bounce model. It is found that the oscillatory behavior of three types of microquasars can also be explained by particle oscillations generated in BBT theory, providing evidence for exploring the existence of wormholes, under the assumptions of parametric resonance and forced resonance.

\end{abstract}


\maketitle

\section{$\text{Introduction}$}

It is widely known that General Relativity (GR) predicts the existence of black holes (BH). In recent years, the study of BH physics has made significant progress, including the discovery of gravitational waves \cite{1} and the imaging of black hole shadows \cite{2,3}. These observational findings either indirectly or directly confirm the predictions  of BH in the universe. However, the predictions of GR regarding BH as being subject to inevitable spacetime singularities result in the eventual breakdown of classical physical laws. Although people have hoped to resolve this issue within the framework of quantum gravity, a reliable theory of quantum gravity remains elusive as of today. Physicists have thus endeavored to tackle this problem through diverse approaches, suggesting notions such as regular black holes \cite{4,5,6,7,8,9,10,11,12,rbh-1,rbh-2,rbh-3,rbh-4,rbh-5,rbh-6,rbh-7,rbh-8} and singularity-free gravitational collapse models \cite{13,14,15,16,17,fgc1}.

The idea of regular BH was initially introduced  by Bardeen in 1968 \cite{4}. Simpson and Visser proposed a spacetime metric, known as the black bounce \cite{18}, which built upon this idea. By introducing a length scale parameter $l$ to regularize central singularities, this metric offers a comprehensive characterization of various objects including Schwarzschild solution, regular BH, and traversable wormholes. It provides a straightforward method for demonstrating the impacts of quantum gravity \cite{19}. Numerous authors have investigated the physical characteristics of the black bounce metric and its varieties, encompassing various topics such as quasi-periodic oscillations (QPOs), gravitational lensing effects, quasi-normal mode frequencies, shadows, and accretion disks \cite{19,20,21,22,23,24,25,26,27,28,29,30,31}. However, research has uncovered inconsistencies between the black bounce model and certain observations \cite{19}.

In addition to BH, wormholes are another significant theoretical prediction of GR. However, in General Relativity, the formation of a wormhole requires the existence of exotic matter that violates the null energy condition \cite{32,33,34}. Exotic matter is commonly rationalized as quantum fields possessing negative energy density within the framework of quantum gravity physics. Although there is currently no astronomical observation that confirms the existence of wormholes, recent research in wormhole physics has been dedicated to exploring observable signals, which are based on theoretical studies \cite{35,36,37,wh-frt,wh-fq}. Several studies suggest that visible indications nearby wormholes might comprise induced gravitational lensing \cite{38,39,40,41}, shadows \cite{42,43,44,45}, and accretion disk radiation \cite{46,47}. The exploration of various effects induced by BH and wormholes offers a theoretical foundation for differentiating various types of celestial objects in observations, while also enabling a comprehensive analysis of the central objects' properties. Reference \cite{38} differentiates between Schwarzschild BH and Ellis wormholes through an analysis of Einstein rings and gravitational lensing. Reference \cite{48} employs the kinematic displacement of photon frequencies to differentiate between BH and wormholes. Reference \cite{49} examines the variation in accretion mass among rotating wormholes and Kerr BH with equivalent mass and accretion rate, revealing that the emission spectra from accretion disks can be utilized to discern the geometric shape of wormholes. In this paper, we aim to explore the distinctive features induced by BH and wormholes in the context of black bounce-type (BBT) geometry, utilizing the high-frequency quasi-periodic oscillations (HFQPOs) method. Our aim is to establish a theoretical framework to account for potential observational disparities between the two, and to facilitate the exploration of various compact celestial bodies and their discernment in observations.

Quasi-periodic oscillations (QPOs), as one of the powerful tools for testing gravitational theories, have been extensively studied by researchers \cite{50,51,52,53,54,55,56,57,QPO-1,QPO-2}. QPOs correspond to peaks observed in the radio-to-X-ray bands of the electromagnetic spectrum emitted by compact objects, as stated in reference \cite{58}. Based on their observed oscillation frequencies, these oscillations are categorized into low-frequency QPOs and high-frequency QPOs. By analyzing the spectra of QPOs \cite{48,58,59,60,61}, scientists can extract certain physical information about the central celestial object. Although the specific causes of QPOs are not fully understood, it is often believed that they are induced by precession and resonance phenomena related to the effects of GR \cite{62,63,64}. In this paper, we apply observations of microquasars to constrain and explore the BBT theoretical model, and investigate the potential physical mechanisms underlying the generation of QPOs.

The structure of this paper is as follows. Section II briefly introduces the BBT theory \cite{25}, and shows the action for the BBT spacetime. In section III, the stable circular orbit regions and the innermost stable circular orbit (ISCO) are investigated for various celestial bodies in BBT spacetime. Section IV centers on particles that experience oscillatory motion around the central celestial object on stable circular orbits, and we compute their inherent radial and azimuthal epicyclic angular frequencies. Furthermore, utilizing models such as parametric resonance and forced resonance in HFQPOs, we conduct an analysis of the resonance locations for various types of celestial bodies in BBT spacetime, under different ratios of intrinsic radial and azimuthal epicyclic angular frequencies. In  section V of this paper, we employ two different resonance models to fit observational data and impose constraints on the parameter $a$ in the black bounce-type spacetime. In addition,  we explore the feasibility of examining various celestial bodies in BBT by using three distinct sets of microquasar oscillation data, and examine the potential physical mechanisms that give rise to HFQPOs. The sixth section concludes the paper.

\section{$\text{A black bounce-type metric}$}

Considering a static spherically symmetric spacetime geometry, its metric can be expressed as \cite{25}:
\begin{equation}
dS^2=-A(x) dt^2+B(x)dx^2+r^2(x) d \Omega^2,\label{1}
\end{equation}
where $A(x)$, $B(x)$ and $r(x)$ are three unspecified functions, the domain of the radial coordinate is $x \in(-\infty,+\infty)$, and $d \Omega^2=d \theta^2+\sin ^2 \theta d \phi^2$ describes the line element of a two-dimensional sphere. For the BBT geometry that we are investigating, proposed by Lobo et al. in reference  \cite{25}, the metric functions can be written as:
\begin{equation}
A(x)=B^{-1}(x)=1-\frac{2 Mx^2}{\left(x^2+a^2\right)^{3 / 2}} ;~~ r^2(x)=x^2+a^2, \label{2}
\end{equation}
where $a$ and $M$ are two constant parameters. Based on the Fan-Wang mass function \cite{Fan-Wang-mass-prd}, Ref.\cite{25} indicates that solution (\ref{2}) can be as a special case appeared in a class of general metric function: $A(x)=B^{-1}(x)=1-\frac{2m(x)}{\Sigma(x)}$, with $m(x)=\frac{M\Sigma(x)x^{k}}{(x^{2n}+a^{2n})^{(k+1)/(2n)}}$ and $n=1$ and $k=2$. For taking other values of constant parameters (e.g. $n=1$ and $k=0$), expressions (\ref{2}) will reduce to black bounce model \cite{18}: $A(x)=B^{-1}(x)=1-\frac{2 M}{\left(x^2+a^2\right)^{1 / 2}}$. It is important to provide an explicit form for the action of system that corresponds to solution (\ref{2}) of the gravitational field equation, which can uplift the status of BBT metric from ad-hoc mathematical model to an exact solution of gravitational theory. Following the method in Ref.\cite{BB-action-prd}, the BBT solution (\ref{1}) with signature $(-,+,+,+)$ can be given by the following action \cite{BBT-action-prd}:

\begin{equation}
S=\int d^4 x \sqrt{-g}\left[R-2 \kappa^2\left(g^{\mu \nu} \partial_\mu \phi \partial_\nu \phi+V(\phi)\right)-2 \kappa^2 L(F)\right],  \label{action-BBT}
\end{equation}
with
\begin{equation}
V(\phi)=\frac{4 M \cos ^5(\phi \kappa)\left(7 \sin ^2(\phi \kappa)-8 \cos ^2(\phi \kappa)\right)}{35 \kappa^2\left|q^3\right|}, \label{BBT-v}
\end{equation}
\begin{equation}
L(F)=\frac{4 \sqrt[4]{2} F^{5 / 4} M(91-75 \sqrt{2 F} q)}{35 \kappa^2 \sqrt{|q|}}.  \label{BBT-LF}
\end{equation}
Here the parameter $a=q$ is the magnetic charge, and $R$ is the Ricci scalar, $g$ is the determinant of the metric, $\phi$ is a non-canonical phantom field, $\kappa^2=8\pi G$ with the gravitational constant $G$, $V(\phi)$ is the potential of $\phi$, $L(F)$ is the Lagrangian for a nonlinear electromagnetic field $F_{\mu\nu}$ with $F=F_{\mu\nu}F^{\mu\nu}/4=q^2/2(a^2+x^2)^2$.
Obviously, the action (\ref{action-BBT}) denotes a gravitational system, at which Einstein's gravitational field minimally coupled with a self-interacting phantom scalar field combined with a nonlinear electrodynamics field. It is well known, the phantom field as a famous dark energy candidate with the equation of state $w<-1$, has been wildly applied to interpret the late accelerating expansion of universe. Also, phantom could appear in string theory in the form of negative tension branes, which play an important role in string dualities \cite{BB-action-prd,phantom1,phantom2}. In fact, in the framework of GR, one of the necessary conditions for forming a wormhole is that one needs to introduce an amount of exotic matter that violates the null energy condition \cite{wh-nec}, e.g. the phantom field. A plenty of wormhole solutions with various kinds of phantom matter were proposed \cite{BB-action-prd,phantom-wh1,phantom-wh2,phantom-wh3,phantom-wh4}.

BBT solution has some attractive properties. For example, (I) it is a simple one-parameter extension of the Schwarzschild metric; (II) It is a candidate of regular BH geometry in the framework of GR, then avoiding the singularity of spacetime of BH. In contrast to singular black holes, the BBT metric restores the integrity of spacetime geodesics,  because the area of the two-dimensional sphere $S=4 \pi r^2(0)=4 \pi a^2$ is finite at $x=0$. The bouncing nature of the radial function can be interpreted as a signal of the existence of a wormhole throat, at which point spacetime is divided into two asymptotically flat regions: $x_{-} \in(-\infty, 0), x_{+} \in(0,+\infty)$. Clearly, when $a \rightarrow 0$, the wormhole throat vanishes, and the above metric degenerates into the form of a Schwarzschild BH, i.e., $A(x) \approx 1-2 M / r$. In the asymptotic limits $x \rightarrow$ $\pm \infty$ and $x \rightarrow 0$, metric  (\ref{2}) corresponds to the forms of Schwarzschild solution and de Sitter solution, respectively, ensuring that the curvature scalar does not diverge;
(III) BBT as a simple model and a unified treatment of distinct kinds of geometries, it smoothly interpolates between some typical BHs and traversable WH.
It can be seen that the above static spherically symmetric metric  (\ref{2}) can describe Schwarzschild BH, double-horizon regular BH, extreme BH, and traversable wormholes for different values of parameter $a$. Specifically, when $a=0$ and $M>0$, it is equal to Schwarzschild BH; when $0<a / M<4 \sqrt{3} / 9$, it describes a regular BH with two horizons; when $a/M=4 \sqrt{3} / 9$, it corresponds to an
extreme black hole; and when $a/M>4 \sqrt{3} / 9$, it represents a traversable wormhole \cite{65}. This paper considers the relevant properties of the BBT theoretical model in conjunction with observational data, given the inconsistencies between the black bounce model and certain observational data \cite{19} and the intriguing properties of the spherical BBT spacetime metric mentioned above.

 \section{$\text{Stable circular orbits and ISCOs for different types of celestial bodies in BBT spacetime}$}
In BBT spacetime, the motion of particles follows the following equation \cite{59}:
\begin{equation}
n=\frac{1}{2}\left[-\frac{1}{~A(x)}\left(\frac{\partial S}{\partial t}\right)^2+\frac{1}{~B(x)}\left(\frac{\partial S}{\partial x}\right)^2+\frac{1}{r^2(x)}\left(\frac{\partial S}{\partial \theta}\right)^2+\frac{1}{r^2(x) \sin ^2 \theta}\left(\frac{\partial S}{\partial \phi}\right)^2\right],   \label{3}
\end{equation}
here $S$ is the action function, which can be related with the 4-momentum of  particle: $p_\mu \equiv {\partial S}/{\partial x^\mu}$. $p^\mu$ is defined as $p^\mu=d x^\mu / d \lambda$  with the affine parameter $\lambda$. For $n=0$, Eq.(\ref{3}) corresponds to the motion of massless particles (e.g., photons), while $n=-1 / 2$ corresponds to the case of massive particles. We set $\theta=\pi / 2$ (the equatorial plane) without any loss of generality. In the BBT geometry, a thin accretion flow is assumed to move along a Keplerian stable circular orbit, which in this case is represented by $\dot{\theta}=0$. Here $\theta$ represents the angular coordinate and the "dot" denotes the derivative with respect to proper time. The timelike geodesic equations for massive particles are expressed as follows:
\begin{equation}
\dot{t}=\frac{dt}{d \lambda}=\frac{E}{A(x)} \label{4}
\end{equation}
 \begin{equation}
\frac{d x}{d \lambda}=\sqrt{E^2-A(x)\left(\frac{J^2}{r^2(x)}+1\right)} \label{5}
\end{equation}
\begin{equation}
\dot{\phi}=\frac{d \phi}{d \lambda}=\frac{J}{r^2(x)}. \label{6}
\end{equation}
Here,  $E$ and $J$ stand for energy and angular momentum, respectively. By utilizing equation  (\ref{5}), we derive the effective potential $V_{\text {eff }}$ for the movement of massive particles on the equatorial plane:
\begin{equation}
V_{e f f}=A(x)\left(1+\frac{J^2}{r^2(x)}\right). \label{7}
\end{equation}
Using the circular orbit condition $dV_{\text {eff }} / dx=0$, we obtain:
\begin{equation}
2 a^4 M-M x^4+J^2 x^2\left(-3 M+\sqrt{a^2+x^2}\right)+a^2\left(M x^2+J^2\left(2 M+\sqrt{a^2+x^2}\right)\right)=0. \label{8}
\end{equation}
In general, people can derive the radial coordinate position of a circular orbit based on equation  (\ref{8}). However, for the BBT metric under consideration, we cannot directly obtain an analytical expression for the circular orbit position using equation  (\ref{8}). Through observation, it is evident that equation  (\ref{8}) is quadratic in relation to a particular angular momentum $J$, thus enabling the determination of circular orbits through the following relationship:
\begin{equation}
J_{c \pm}= \pm \sqrt{\frac{M\left(-2 a^4-a^2 x^2+x^4\right)}{x^2\left(-3 M+\sqrt{a^2+x^2}\right)+a^2\left(2 M+\sqrt{a^2+x^2}\right)}}.  \label{9}
\end{equation}
$J_{c+}, ~J_{c-}$ represents the angular momentum in two possible directions when particles perform circular motion around the central celestial object in the equatorial plane. And the energy of particles on circular orbits is:
\begin{equation}
E=\frac{-2 M x^2+\left(a^2+x^2\right)^{3 / 2}}{\sqrt{\left(a^2+x^2\right)^{3 / 2}\left(x^2\left(-3 M+\sqrt{a^2+x^2}\right)+a^2\left(2 M+\sqrt{a^2+x^2}\right)\right)}}. \label{9.5}
\end{equation}
In the context of BBT geometry, it is clear that the angular momentum  (\ref{9}) is symmetric with respect to the radial coordinate $x$. For the purpose of this paper, we have chosen the case of $x \geq 0$ for discussion. In order to provide significance to equation  (\ref{9}), it is necessary to impose limitations on the domain of the radial coordinate $x$ and establish the area where particles have the ability to execute circular orbit motion. Calculations reveal that when $a \geq 4 \sqrt{3} M / 9$, the circular orbit interval exists within:
\begin{equation}
|x| \geq \sqrt{2} a. \label{10}
\end{equation}
And when $a<4 \sqrt{3} M / 9$, the circular orbit region is confined to:
\begin{equation}
|x|>\sqrt{-a^2+3 M^2-\frac{10 \sqrt[3]{2} a^2 M^2+9 \sqrt[3]{2} M^4}{p}+\frac{p}{\sqrt[3]{2}}},  \label{11}
\end{equation}
where $p=\left(25 a^4 M^2-90 a^2 M^4+54 M^6+5 \sqrt{5} \sqrt{5 a^8 M^4-4 a^6 M^6}\right)^{1 / 3}$.

\begin{figure}[ht]
  \includegraphics[width=8cm]{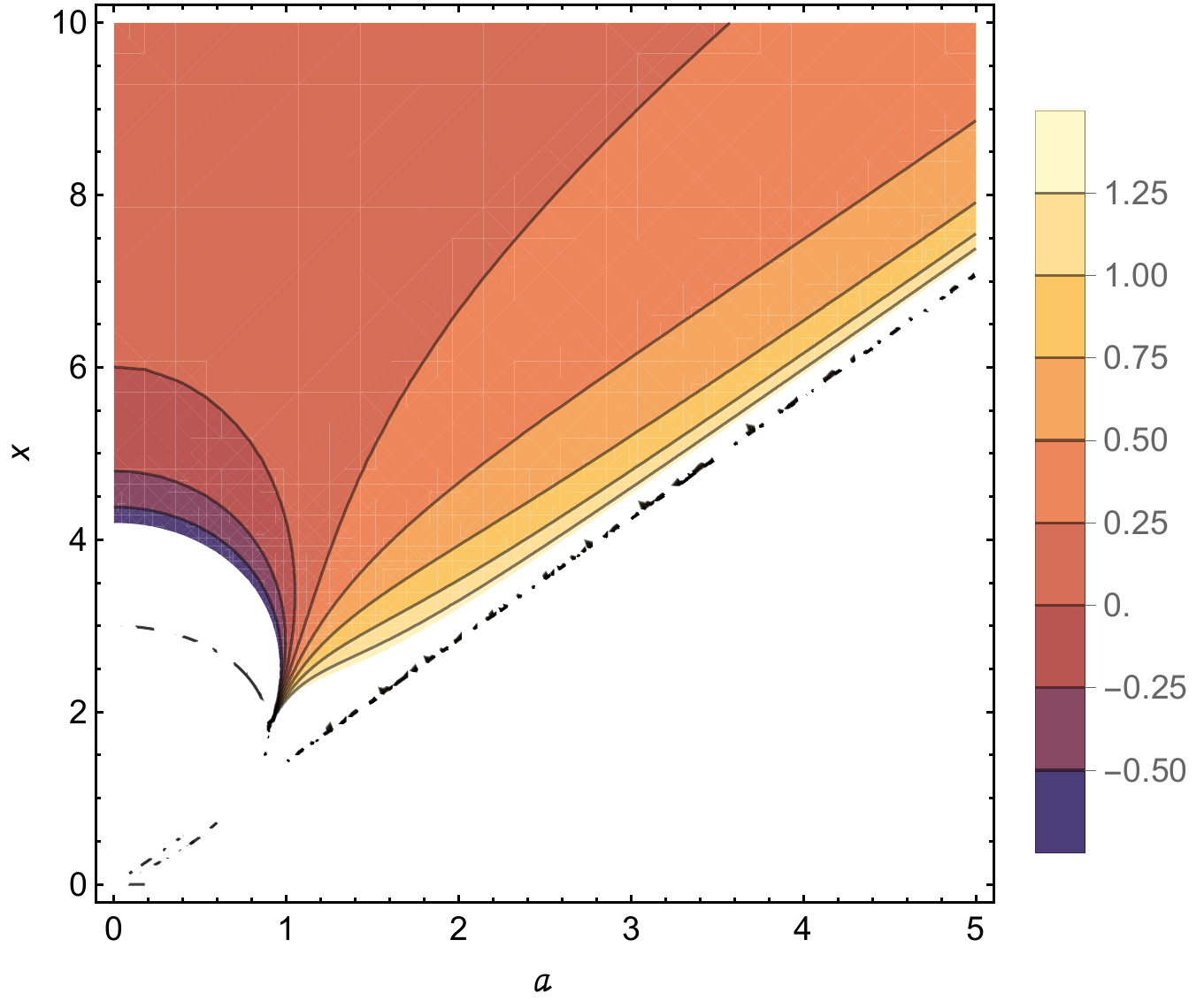}
  \caption{Variations of ISCO $\left(dJ_{c+} / dx=0\right)$ and stable circular orbit positions $\left(d J_{c+} / d x \geq 0\right)$ relative to the parameter $a$ for different types of celestial bodies in BBT spacetime.}\label{fig:1}
\end{figure}

Next, we analyze the stability of circular orbits. Clearly, when $d J_{c+} / d x \geq 0$, it corresponds to stable circular orbits where the angular momentum $J_{c+}$ has a local extremum, namely $dJ_{c+} / dx=0$ corresponding to the ISCO. ISCO serves as the inner boundary of the accretion disk and the starting point of electromagnetic radiation, making it crucial in the study of accretion disks around compact objects \cite{66,67,68,69,70}. For the BBT model, we derive using $dJ_{c+} / dx \geq 0$, the following:
\begin{equation}
\frac{4 a^6 x+x^7-6 x^5 \sqrt{a^2+x^2}+a^4\left(9 x^3-16 x \sqrt{a^2+x^2}\right)+a^2\left(6 x^5+8 x^3 \sqrt{a^2+x^2}\right)}{2 \sqrt{a^2+x^2} \sqrt{-2 a^4-a^2 x^2+x^4}\left(x^2\left(-3+\sqrt{a^2+x^2}\right)+a^2\left(2+\sqrt{a^2+x^2}\right)\right)^{3 / 2}} \geq 0.   \label{12}
\end{equation}
Equation (\ref{12}) indicates that the position of stable circular orbits $x$ varies with different values of $a$, which corresponds to different types of celestial bodies. We establish the relationship between them through numerical calculations (as shown in Figure \ref{fig:1}). Without loss of generality, we set $M=1$ in this paper. From Figure \ref{fig:1}, it can be observed that: when $a \leq 4 \sqrt{3} / 9$, there exists an ISCO around celestial bodies (Schwarzschild BH, regular BH, extremal BH). When $a>4 \sqrt{3} / 9$, celestial bodies (traversable wormholes) have two ISCOs (for $4 \sqrt{3} / 9< a \lesssim 1.050$), or one ISCO (for $a > 1.050$). It should be noted that in the case of two ISCOs, there is an unstable circular orbit region between them, where there is a "vacuum" annular region between the accreting matter around celestial bodies, similar in nature to the Janis-Newman-Winicour spacetime \cite{71}.

\begin{figure}[htbp]
  \includegraphics[width=8cm]{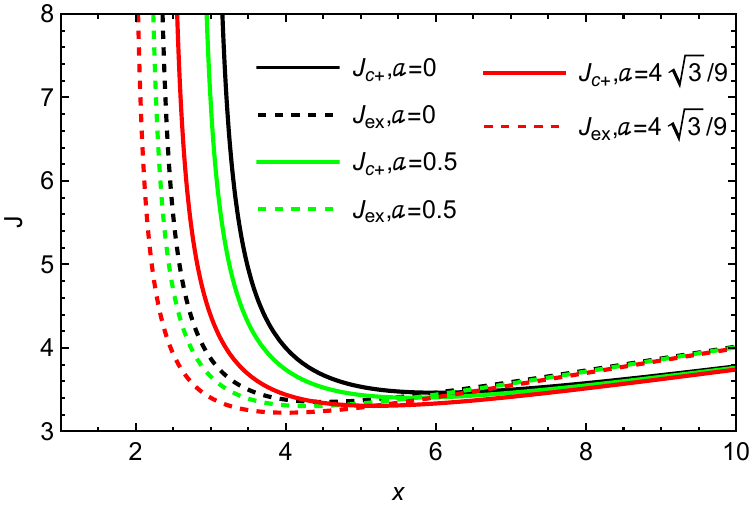}
  \includegraphics[width=8cm]{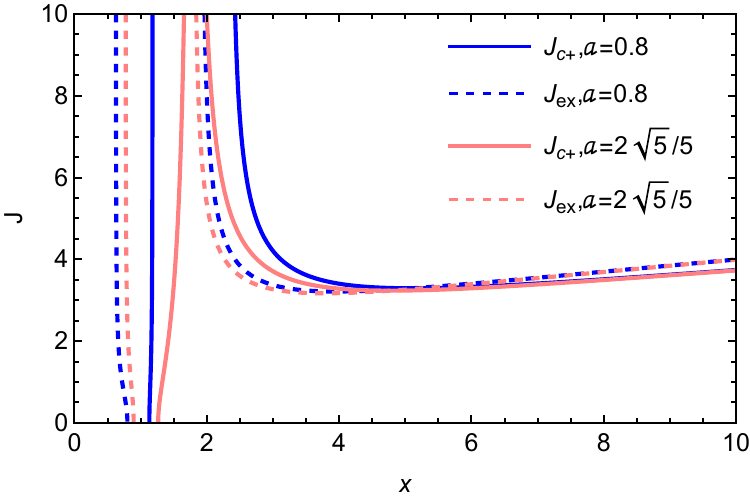}
   \includegraphics[width=8cm]{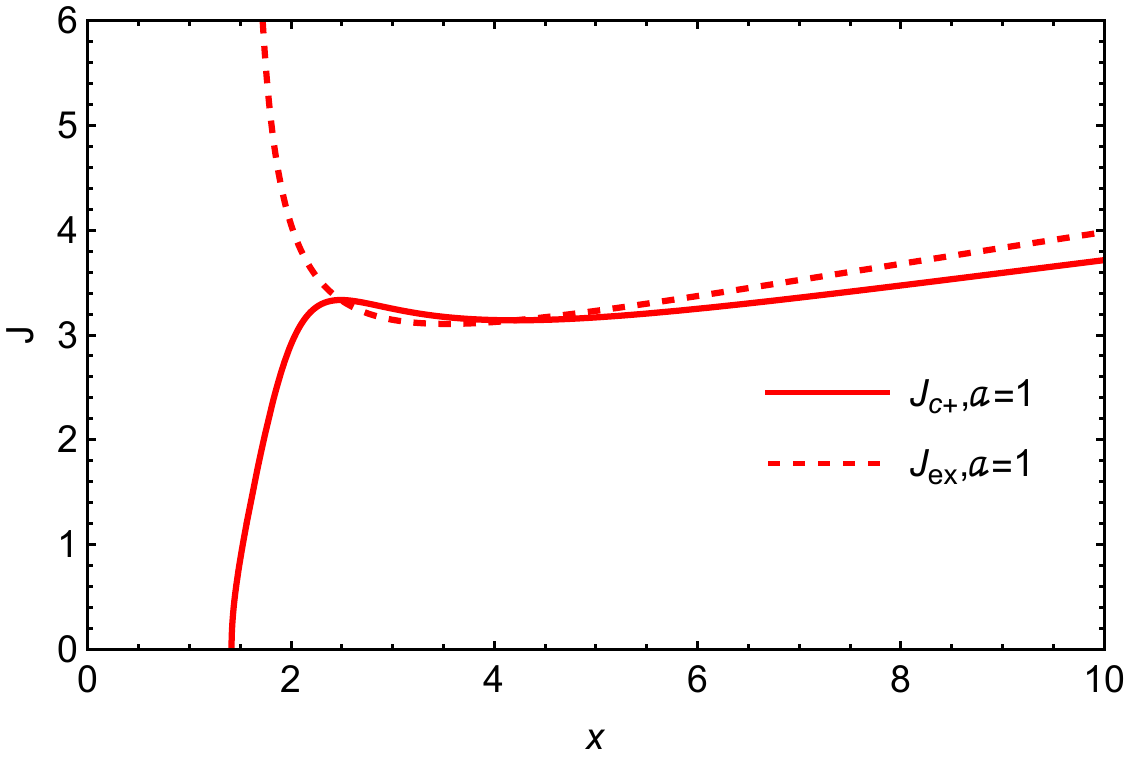}
  \includegraphics[width=8cm]{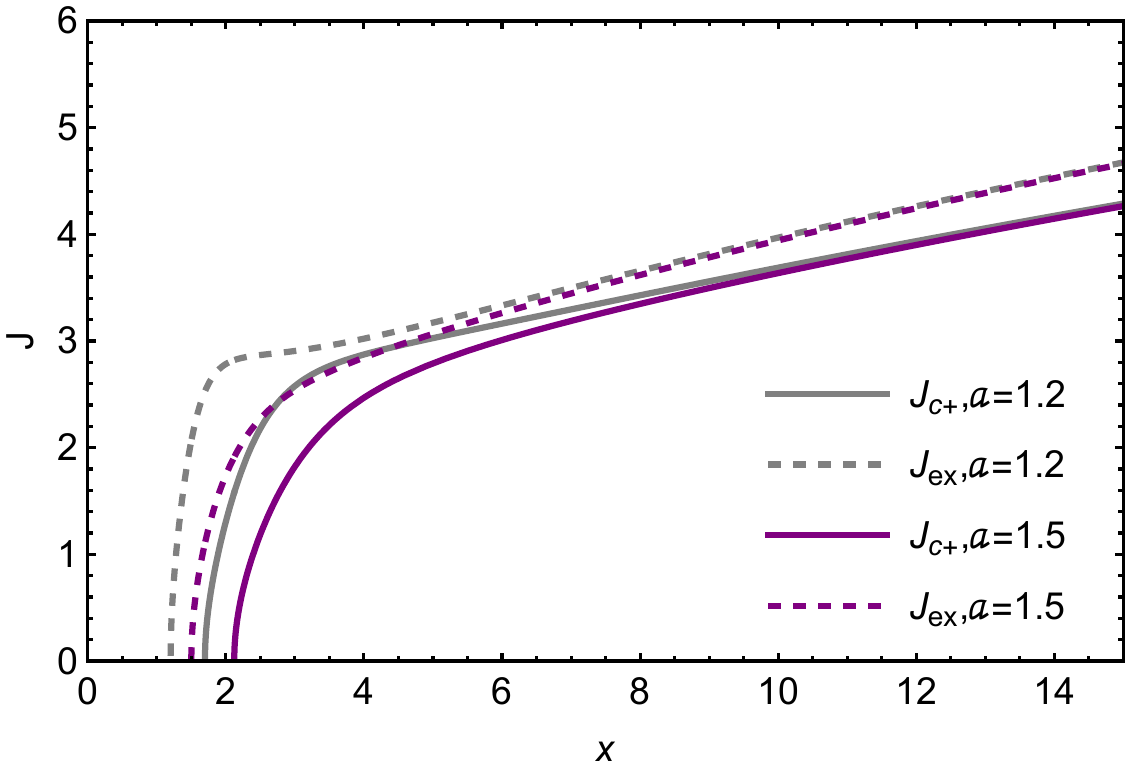}
    \caption{ Variation curves of $J_{c+}$ and $J_{ex}$ relative to $x$ for various types of celestial bodies (with varying values of $a$), where solid lines represent $J_{c+}$ and dashed lines represent $J_{ex}$.}
    \label{fig:2}
\end{figure}

Furthermore, Figure \ref{fig:1} reveals that the expression $dJ_{c+} / dx>0$ is consistently held when the value of $a$ is  larger (e.g., $a \gtrsim 1.050$ ), thereby indicating our inability to determine the position of ISCO through calculation $dJ_{c+} / dx=0$. Since all circular orbits that correspond to $dJ_{c+} / dx>0$ are stable, we can calculate the position of ISCO by intersecting $J_{c+}$ with the $x$-axis. Figure  \ref{fig:2} (bottom right) illustrates the variation of $J_{c+}$ relative to $x$ when $a \gtrsim 1.050$. For instance, consider $a=1.2$ and 1.5. In addition, to offer a more intuitive depiction of the ISCO properties corresponding to different types of celestial bodies, we also plot $J_{c+}$ and $J_{ex}$ in Figure \ref{fig:2} for specific values of $a$ (the intersection of the two represents ISCO). The expression for $J_{ex}$ can be derived from $d^2 V_{\text {eff }} / dx^2=0$:
\begin{equation}
J_{e x}=\frac{\sqrt{-2 a^4-3 a^2 x^2+5 x^4}}{\sqrt{a^2\left(2+\sqrt{a^2+x^2}\right)+x^2\left(-9+4 \sqrt{a^2+x^2}\right)}}. \label{13}
\end{equation}

From Figure \ref{fig:2} (top left), it becomes evident that for Schwarzschild BH ($a=0$), regular BH (e.g., considering $a=0.5$), extreme BH ($a=$ $4 \sqrt{3} / 9$), there exists a single intersection point in their respective $J_{c+}$ versus $J_{ex}$ graphs. If we label the position of this intersection point as $x_{\text {ISCO}}$, then the regions corresponding to stable circular orbits are represented as $x \geq x_{\text {ISCO}}$, while the unstable circular orbit regions are $x<x_{\text {ISCO}}$.

For other plots in Figure \ref{fig:2}, we show the stable circular orbits for wormholes. Concretely, (1) for the case of a traversable wormhole with two or one photon sphere ($4 \sqrt{3} / 9<a\leq 2\sqrt{5} / 5$),  e.g., taking $a=0.8, 2\sqrt{5} / 5$, as observed in Figure \ref{fig:2} (top right), the plot is divided into two segments, which means there exist two regions of stable circular orbits. For the left part of this picture, we can derive the position of the ISCO using the following general relation: $x=\sqrt{2} a$, which is located at the intersection of the solid line and the horizontal axis. And for the right part, the intersection points of $J_{c+}$ and $J_{ex}$ represent the position: $x_{\text {ISCO}}$. Then $x>x_{\text {ISCO}}$ describes stable circular orbits, while $x<x_{\text {ISCO}}$ denotes unstable circular orbits;
(2) For the case of traversable wormholes with a single photon sphere ($2 \sqrt{5} / 5<a\lesssim 1.050$),  the stable circular orbits are also divided into two segments when we set $a=1$ as an example, as shown in Figure \ref{fig:2} (bottom left). But unlike the top-right case, the curves of $J_{c+}$ and $J_{ex}$ are continuous for bottom-left picture. The stable circular orbits correspond to the position intervals of $\sqrt{2} \leq x \lesssim 2.491$  and $x \gtrsim 4.203$, respectively. The interval region between $J_{c+}$ and $J_{ex}$ intersections ($2.491 \lesssim x \lesssim 4.203$) corresponds to unstable circular orbits;
(3) When $a$ is taken the larger values, such as  $a=1.2$ or $ 1.5$, the stable circular orbit region becomes continuous, and the position of ISCO is given by the intersection of $J_{c+}$ and the $x$-axis: $x=6 \sqrt{2} / 5$ and $3 \sqrt{2} / 2$.

\begin{figure}[htbp]
  \includegraphics[width=8cm]{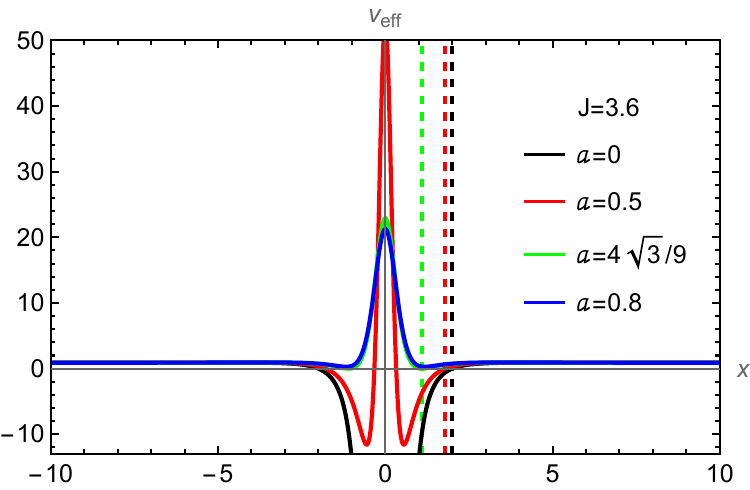}
  \includegraphics[width=8cm]{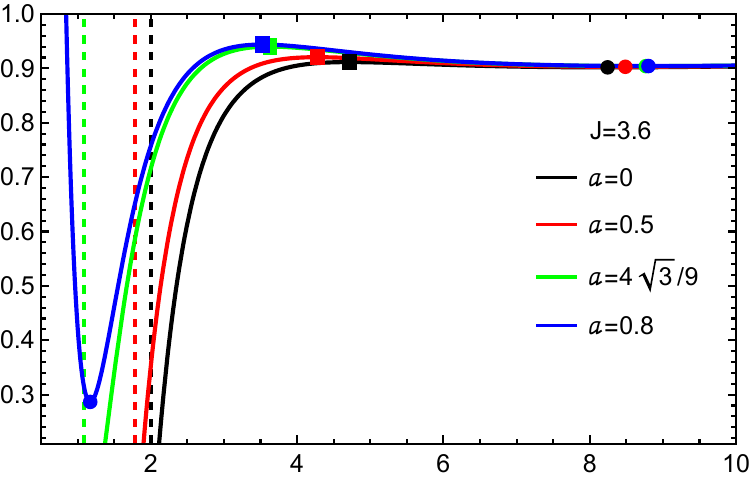}
  \includegraphics[width=8cm]{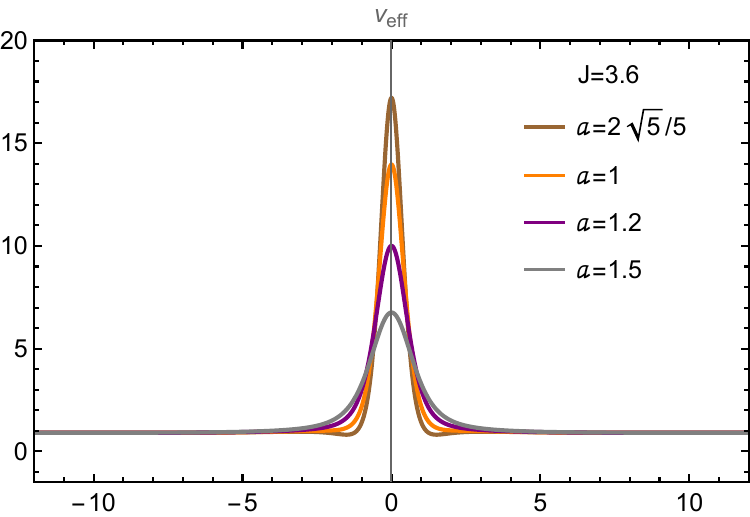}
  \includegraphics[width=8cm]{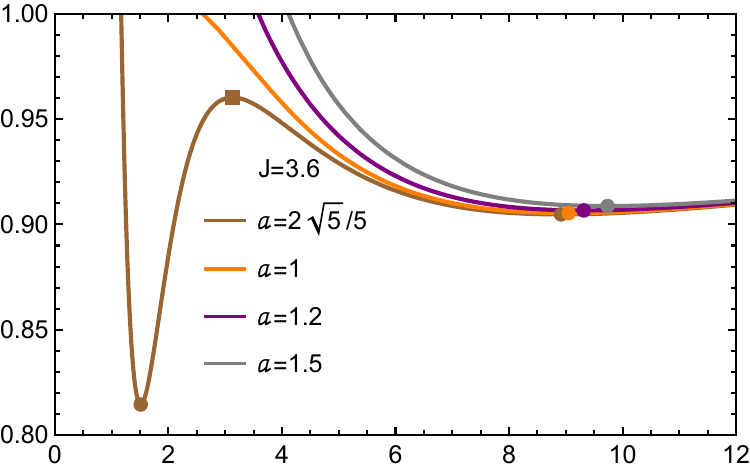}
    \caption{The left pictures shows the effective potential for test particles with $J=3.6$, where $a=0$ corresponds to the Schwarzschild BH, $0<a<4 \sqrt{3} / 9$ describes the regular BH, $a=4 \sqrt{3} / 9$ corresponds to the extremal BH, and $a>4 \sqrt{3} / 9$ represents the traversable wormhole. The positions of the event horizons for $a$=0, $a=0.5$, and $a \leq 4 \sqrt{3} / 9$ are represented by black, red, and green dashed lines, respectively. The right sides are the zoomed-in view of the left pictures, where the dots represent the positions of stable circular orbits, and the squares represent the positions of  unstable circular orbits for various celestial bodies.}
    \label{fig:3}
\end{figure}

In order to demonstrate properties of the effective potential associated with various types of celestial bodies, we utilize equation (\ref{7}) to graph the variation curve of the effective potential with respect to the radial coordinate in Figure \ref{fig:3} (left). In order to distinguish the effective potential images of various celestial bodies, a constant parameter $J=3.6$ is designated. Figure \ref{fig:3} (right) presents a close-up and enlarged image of the effective potential that is specifically targeted. In Figure \ref{fig:3} (top), the positions of the event horizons for Schwarzschild BH ($a=0$), regular BH ($a=0.5$), extreme BH ($a=4 \sqrt{3} / 9$) are represented by black, red, and green dashed lines, respectively. When $a>4 \sqrt{3} / 9$, the BBT spacetime describes the traversable wormholes, where the event horizon is not present. Furthermore, in the zoomed-in view, we use circular and square markers to indicate the positions of stable and unstable circular orbits located outside the event horizons for various celestial bodies, respectively.

 Notably, assuming that a particle with an angular momentum of $J=3.6$, it could own  stable circular orbits for all cases of $a$ considered in this paper. Furthermore, consider the particle coming from infinity, there exist unstable circular orbits in the cases of  $a\leq 2 \sqrt{5} / 5$. For the case of unstable circular orbits, if $a\leq 4 \sqrt{3} / 9$, an inward perturbation causes the particle to fall into the black hole and be captured, while an outward perturbation results in the particle flying off to infinity; If $a> 4 \sqrt{3} / 9$ (e.g. $a=0.8$, $a=2 \sqrt{5} / 5$), an outward perturbation likewise causes the particle to fly off to infinity, but an inward perturbation could not make the particle to fall into the wormhole. In contrast, when
${d r}/{d \lambda}=0$, it will return.

\section{$\text{Resonance frequency and resonance position of particles around different types of celestial bodies in BBT}$}

\subsection{{Angular frequency of oscillating particles}}
In this section, we explore the frequency of oscillation of test particles around various celestial bodies in BBT spacetime, on stable circular orbits. If the moving particle assumes a slight deviation from the minimum of the effective potential, it follows that the particle will oscillate on a stable circular orbit, thereby achieving epicyclic motion that is controlled by linear harmonic oscillation. Taking into account $x=x_{c}+\delta x$, where $x_{c}$ represents the radial coordinate at the minimum of the effective potential, and $\delta x$ describes the radial perturbation displacement - it is a small quantity. On the equatorial plane, the transverse displacement in the presence of a small perturbation $\delta \theta$ is represented as $\theta=\pi / 2+\delta \theta$. Under linear perturbations, the equations governing the particle's epicyclic motion around a stable circular orbit in the radial and latitudinal directions may be represented as follows:
\begin{equation}
\delta \ddot{x}+\omega_{x}^2 \delta x=0, \quad \delta \ddot{\theta}+\omega_\theta^2 \delta \theta=0.   \label{14}
\end{equation}
Here, the 'dot' denotes the derivative with respect to the particle's proper time $\tau$, and $\omega_{x}$ (or $\omega_\theta)$ represents the radial (or latitudinal) angular frequency of the particle undergoing oscillatory motion at the circular orbit position. Considering the Hamiltonian:
\begin{equation}
H=H_{dyn }+H_{pot }=\frac{1}{2} ~g^{\alpha \beta} p_\alpha p_\beta+\frac{m^2}{2},    \label{15}
\end{equation}
where
\begin{equation}
H_{d y n}=\frac{1}{2}\left(g^{x x} p_x^2+g^{\theta \theta} p_\theta^2\right),    \label{16}
\end{equation}
\begin{equation}
H_{pot}=\frac{1}{2}\left(g^{t t} E^2+g^{\phi \phi} J^2+1\right),   \label{17}
\end{equation}
correspond to the kinetic and potential energy parts of the Hamiltonian. Here $p_{x}=\frac{\partial S}{\partial x}=\sqrt{\frac{E^2}{A(x)^2}-\frac{1}{A(x)}[\frac{J^2}{r(x)^2}+1]}$, and $p_{\theta}=\frac{\partial S}{\partial \theta}=0$. The angular frequencies $\omega_{x}{ }^2$ and $\omega_\theta{ }^2$ for the radial and latitudinal epicyclic motion, respectively, can be calculated using the following relationships:
\begin{equation}
\omega_{x}^2=\frac{1}{~g_{xx}} \frac{\partial^2 H_{pot}}{\partial x^2},   \label{18}
\end{equation}
\begin{equation}
\omega_\theta^2=\frac{1}{g_{\theta \theta}} \frac{\partial^2 H_{pot}}{\partial \theta^2}.   \label{19}
\end{equation}
For the BBT model studied in this paper, we derive the following:
\begin{equation}
\omega_x=\sqrt{\frac{x^6\left(-6+\sqrt{a^2+x^2}\right)+4 a^4 x^2\left(-4+\sqrt{a^2+x^2}\right)+a^2 x^4\left(8+5 \sqrt{a^2+x^2}\right)}{\left(a^2+x^2\right)^{7 / 2}\left(x^2\left(-3+\sqrt{a^2+x^2}\right)+a^2\left(2+\sqrt{a^2+x^2}\right)\right)}}.    \label{20}
\end{equation}
\begin{equation}
\omega_\theta=\sqrt{\frac{-2 a^2+x^2}{\left(a^2+x^2\right)\left(x^2\left(-3+\sqrt{a^2+x^2}\right)+a^2\left(2+\sqrt{a^2+x^2}\right)\right)}}.   \label{21}
\end{equation}
The angular frequency of particle's orbital (or vertical) motion is represented as:
\begin{equation}
\omega_\phi=\frac{d \phi}{d t}=\frac{J}{g_{\phi \phi}}.   \label{22}
\end{equation}
Clearly, in spherically symmetric spacetimes, we have $\omega_\theta=\omega_\phi$. The primary sources of the QPO phenomenon are considered to be orbital precession and epicyclic motion. Models, such as orbital precession models and resonance models, can be constructed to investigate the behavior of QPOs in celestial bodies. Resonant behavior frequently manifests in accretion disks, enabling researchers to glean valuable insights about the central object and its associated accretion disk by examining QPO phenomena occurring around various celestial bodies. This includes possible excitation of resonance modes, the locations where resonance occurs, and peak frequencies, among other factors \cite{71}.

\begin{figure}[htbp]
  \includegraphics[width=8cm]{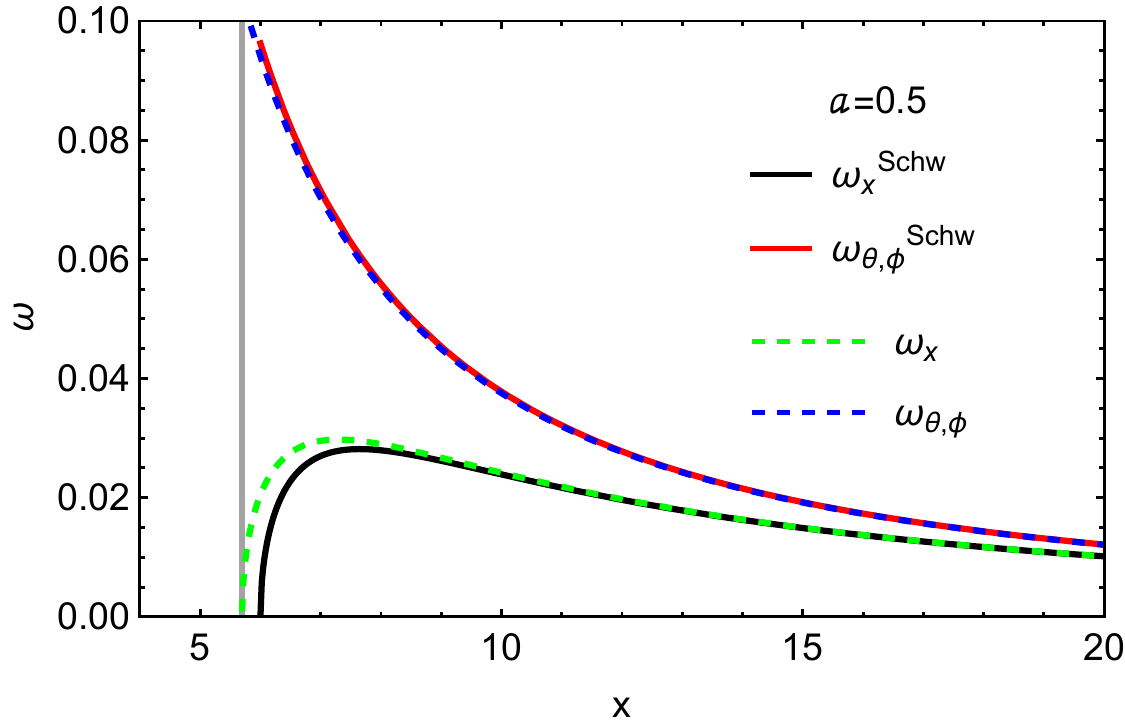}
  \includegraphics[width=8cm]{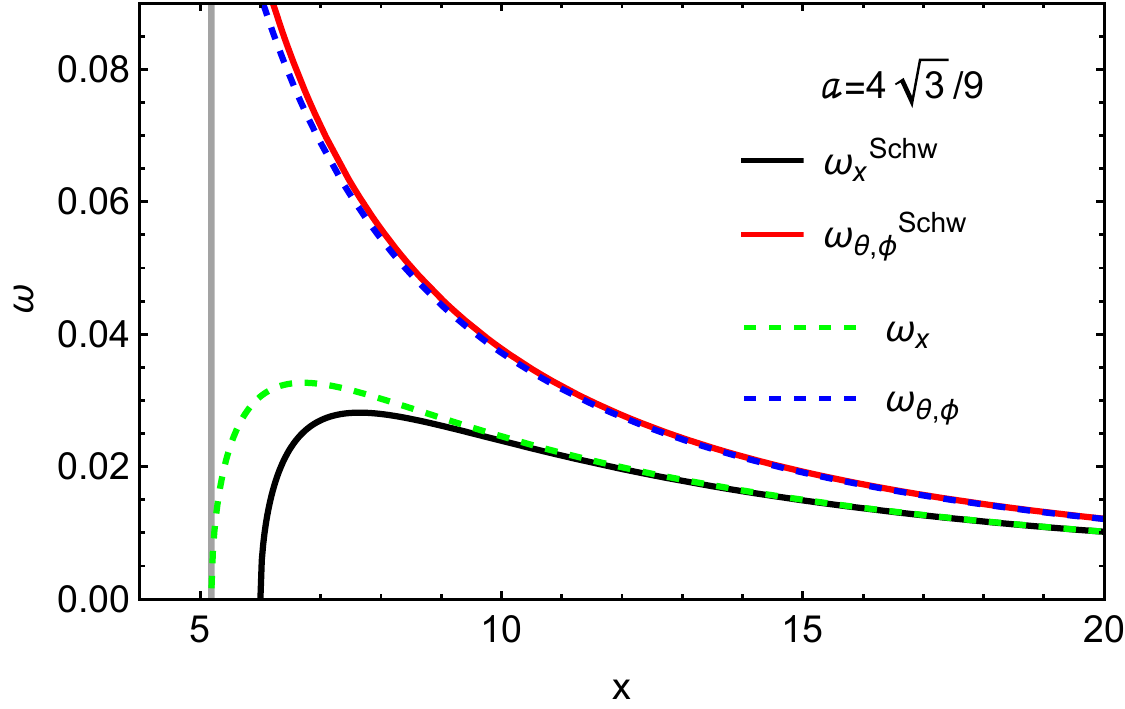}
  \includegraphics[width=8cm]{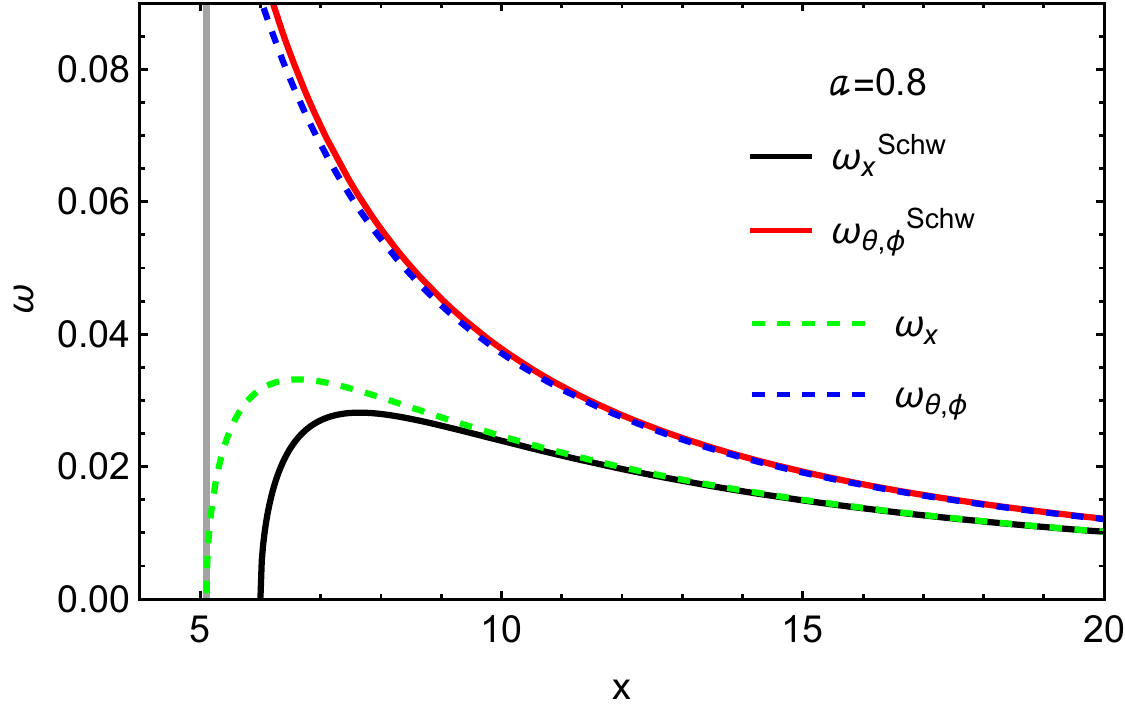}
  \includegraphics[width=8cm]{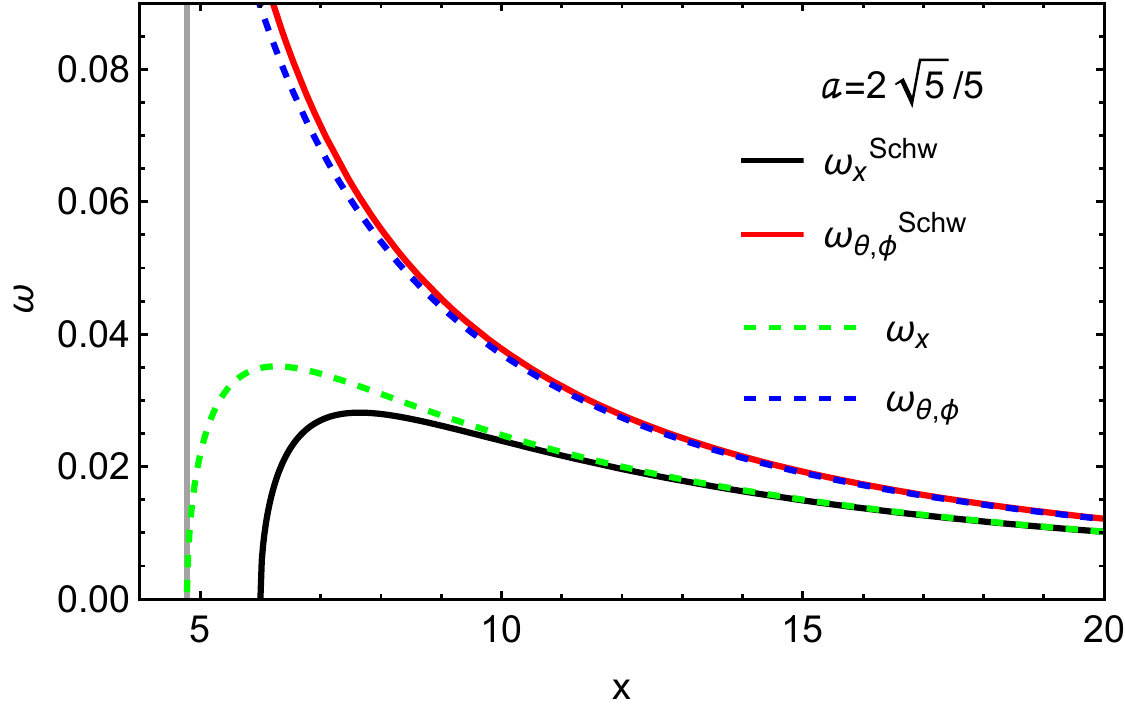}
  \includegraphics[width=8cm]{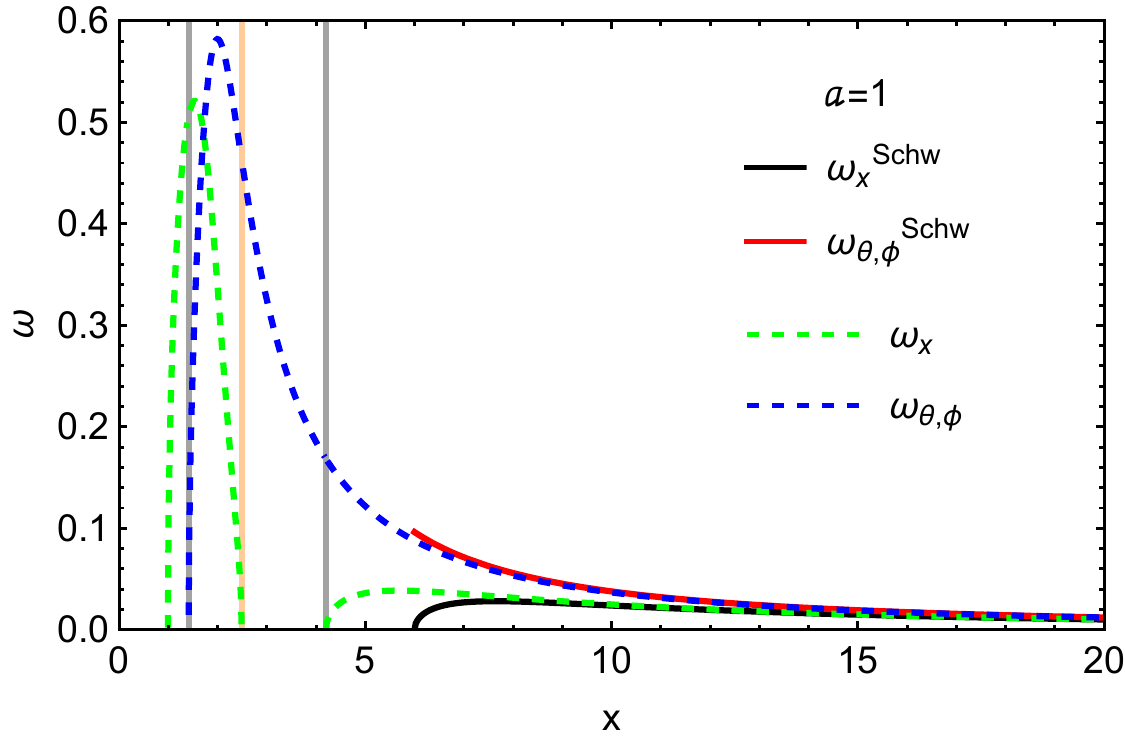}
  \includegraphics[width=8cm]{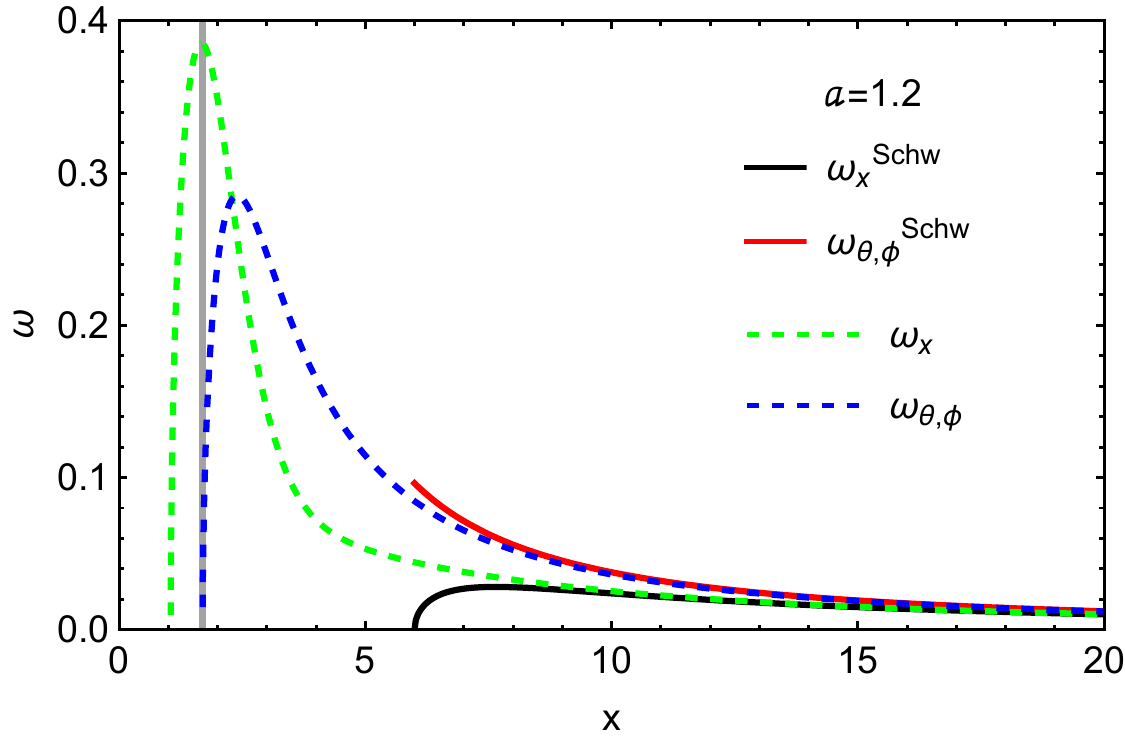}
  \includegraphics[width=8cm]{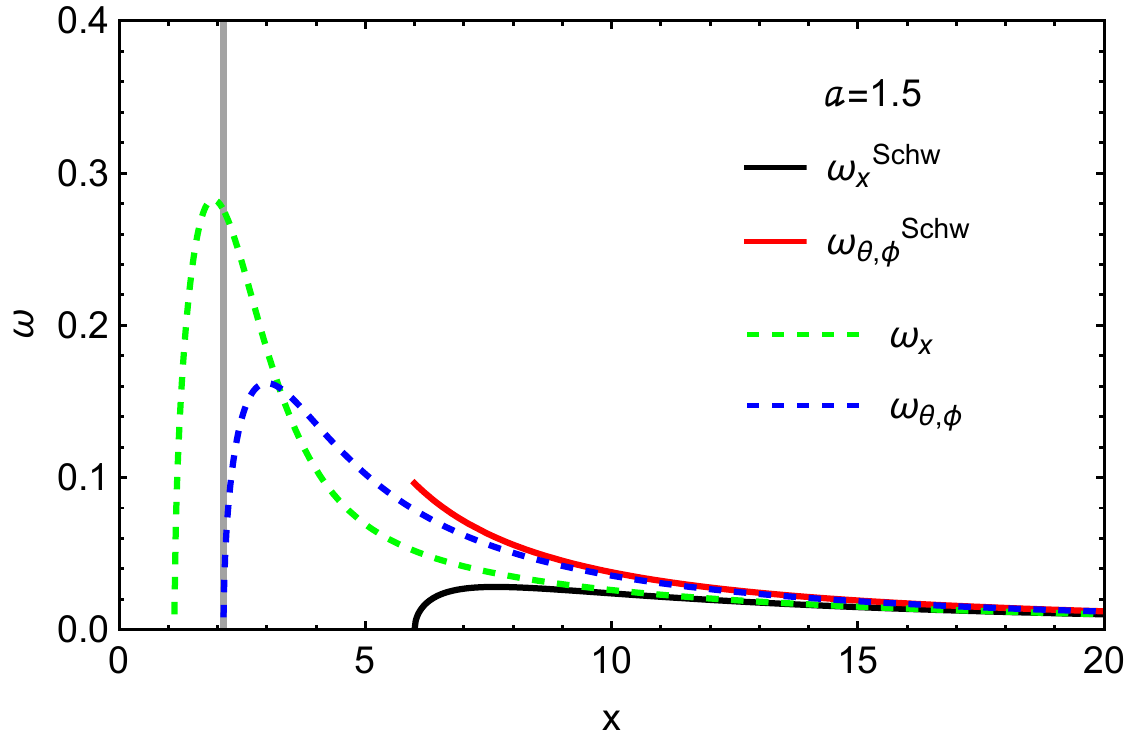}
    \caption{Variation of the particle's radial and latitudinal epicyclic motion frequencies relative to the radial coordinate $x$ for different values of $a$ corresponding to different types of celestial bodies. In the figure, the solid black and red lines represent the radial and latitudinal epicyclic motion angular frequencies for the Schwarzschild BH, while the dashed green and blue lines depict the radial and latitudinal epicyclic motion angular frequencies in the BBT geometry. The gray line denotes the location of the ISCO in the BBT geometry, and the orange line represents the position of the innermost unstable circular orbit.}
    \label{fig:4}
\end{figure}

People typically assume that the epicyclic motion may be caused by the motion of accretion flows inside the accretion disk. Considering that the ISCO serves as the inner boundary of the accretion disk, in our study, we consider the physically meaningful range of the radial coprdinate: $x \geq x_{\text {ISCO }}$. For the Schwarzschild black hole, the computed values for $x_{\text {ISCO }}=6$ are obtained. As seen in Figure \ref{fig:4} (first and second rows), for the Schwarzschild BH $(a=0)$, regular BH (e.g., with $a=0.5$ chosen), extremal BH $(a=4 \sqrt{3} / 9)$, traversable wormhole with double photon spheres (e.g., with $a=0.8$ chosen), and traversable wormhole with a single photon sphere $(a=2 \sqrt{5} / 5)$. The latitudinal epicyclic angular frequencies of particles undergoing simple harmonic motion in the $x \geq x_{\text {ISCO }}$ region are always greater than the radial epicyclic angular frequencies. For the cases of $a=0.8$ and $a=2 \sqrt{5} / 5$, we  plot pictures by only choosing the right segment of stable circular orbits, as shown in Fig. \ref{fig:2} (top right). This choice would not change the conclusions in the BBT spacetime presented below. In the BBT spacetime, the trends of $\omega_{x}$ and $\omega_\theta$ Figure \ref{fig:4} (first and second rows), are similar to the Schwarzschild black-hole case, i.e., $\omega_\theta$ monotonically decreases with increasing radial coordinate $x$ and $\omega_{x}$ exhibits $a$ single-peaked structure. However, for the traversable wormhole with a single photon sphere $(a>2 \sqrt{5} / 5)$, as seen in Figure \ref{fig:4} (third and fourth rows), the $\omega_{x}$ and $ \omega_\theta$ patterns in the BBT model are notably different from the Schwarzschild BH case. Contrary to the black-bounce results reported in reference \cite{19}, we observe the presence of $\omega_x \geq \omega_\theta$ in the BBT model. Furthermore, when $a=1$, the resulting accretion disk has a ring-like structure, leading to a more complex shapes for $\omega_{x}$ and $\omega_\theta$ that need to be represented using piecewise functions. In this case, the region of stable circular orbit is $x_{\text {ISCO1 }} \leq x<x_{\text {orange }}$ and $x_{\text {ISCO2 }} \leq x$, while the region of unstable circular orbits corresponds to $x_{\text {orange }} \leq x<x_{\text {ISCO2 }}$. When $a=1.2$ and 1.5, it differs from the conclusions presented for the case $a \leq 2 \sqrt{5} / 5$ shown in Figure \ref{fig:4} (first and second rows): $\omega_{x}$ exhibits a decaying mode with increasing $x$ value, and $\omega_\theta$ has a single-peaked structure. Finally, comparing Figure \ref{fig:4} (first and second rows) and Figure \ref{fig:4} (third and fourth rows), we observe that the case of traversable wormholes with a single photon sphere corresponds to larger angular frequency values.

\subsection{{Study of resonance positions based on the HFQPOs model}}

A wealth of observational evidence suggests that in low-mass $X$-ray binaries (LMXBs) containing black holes, the double peaks of HFQPOs are often observed with a fixed ratio of high-peak and low-peak frequencies, typically in a 3:2 ratio $\left(\nu_{u}: \nu_{l}\right) $ \cite{72}. Speculation exists that the phenomenon may be caused by a resonance, which is produced by an oscillatory mechanism within the accretion disk. In the preceding sections, we examined the characteristics of oscillation frequencies at circular orbits for various types of celestial bodies in the uncoupled scenario, where perturbations $\delta x$ and  $\delta \theta $  were not linked. However, in many specific scenarios, it is often assumed that there may be dissipation, pressure effects, or the influence of forces such as viscosity and magnetic fields inside the accretion disk, as suggested in references \cite{73,74,75}. This requires taking into account the coupling between $\delta x$ and $\delta \theta$, which implies including associated nonlinear terms in the perturbation equations. Due to the existing constraints on the investigation of accretion disk physics, it is a challenging task to offer a universal mathematical equation to characterize the perturbation behavior. A more practical approach is to establish models by taking into account specific physical circumstances in order to discuss the problem. Various theoretical models have been proposed to explain the observed $\mathrm{QPO}$s phenomenon, including the parametric resonance model, the forced resonance model, the Keplerian resonance model, the non-axisymmetric disk oscillation model, and the relativistic precession model \cite{73}. Here, we explore the parametric resonance model and the forced resonance model, which are frequently encountered in the study of black hole physics and epicyclic motion. Considering the perturbation equations:

\begin{equation}
\delta \ddot{x}+\omega_{x}^2 \delta x=\omega_{x}^2 ~F_{x}(\delta x, \delta \theta, \delta \dot{x}, \delta \dot{\theta}),~~ \delta \ddot{\theta}+\omega_\theta^2 \delta \theta=\omega_\theta^2 ~F_\theta(\delta x, \delta \theta, \delta \dot{x}, \delta \dot{\theta}),   \label{23}
\end{equation}
where $F_x$ and $F_\theta$ represent two undetermined functions corresponding to the coupling effects caused by perturbation terms. In the parametric resonance model \cite{76}, it is assumed that $F_x=0, F_\theta=h \delta \theta \delta x$, and $h$ is constant. In this case, equation (\ref{23}) becomes:
\begin{equation}
\delta \ddot{x}+\omega_x^2 \delta x=0,~~  \delta \ddot{\theta}+\omega_\theta^2\left[1+h \cos \left(\omega_x t\right)\right] \delta \theta=0.   \label{24}
\end{equation}
According to equation (\ref{24}), parametric resonance occurs when the following conditions are satisfied:
\begin{equation}
\frac{\omega_x}{\omega_\theta}=\frac{\nu_x}{\nu_\theta}=\frac{2}{n},~~(n=1,2,3 \ldots).  \label{25}
\end{equation}
Here  $\nu_x=\omega_x / 2 \pi, \nu_\theta=\omega_\theta / 2 \pi$, and $n$ denotes positive integers. Clearly, as the resonance parametric $n$ decreases, the resonance phenomenon becomes more pronounced \cite{74}. In the case of a BBT spacetime, when $a \leq 2 \sqrt{5} / 5$, we have $\omega_\theta>$ $\omega_x$, which prevents the lowest-order resonance parameters $(n=1, n=2)$ from being excited. This means that for the central celestial bodies (including BH and wormhole)  corresponding to this situation, the minimum value of the resonance parametric $n$ can only be 3. However, for larger values of $a$ $(a>2 \sqrt{5} / 5)$, because the relationship between the radial and latitudinal epicyclic oscillation frequency values is uncertain (i.e., $\omega_\theta>\omega_x$, $\omega_\theta<\omega_x$, and $\omega_\theta=\omega_x$ can all occur), this suggests that low-order resonance parameters $(n=1, n=2)$ can be excited in such celestial bodies. This is different from what is implied in the case of $a \leq 2 \sqrt{5} / 5$.

\begin{figure}[htbp]
  \includegraphics[width=10cm]{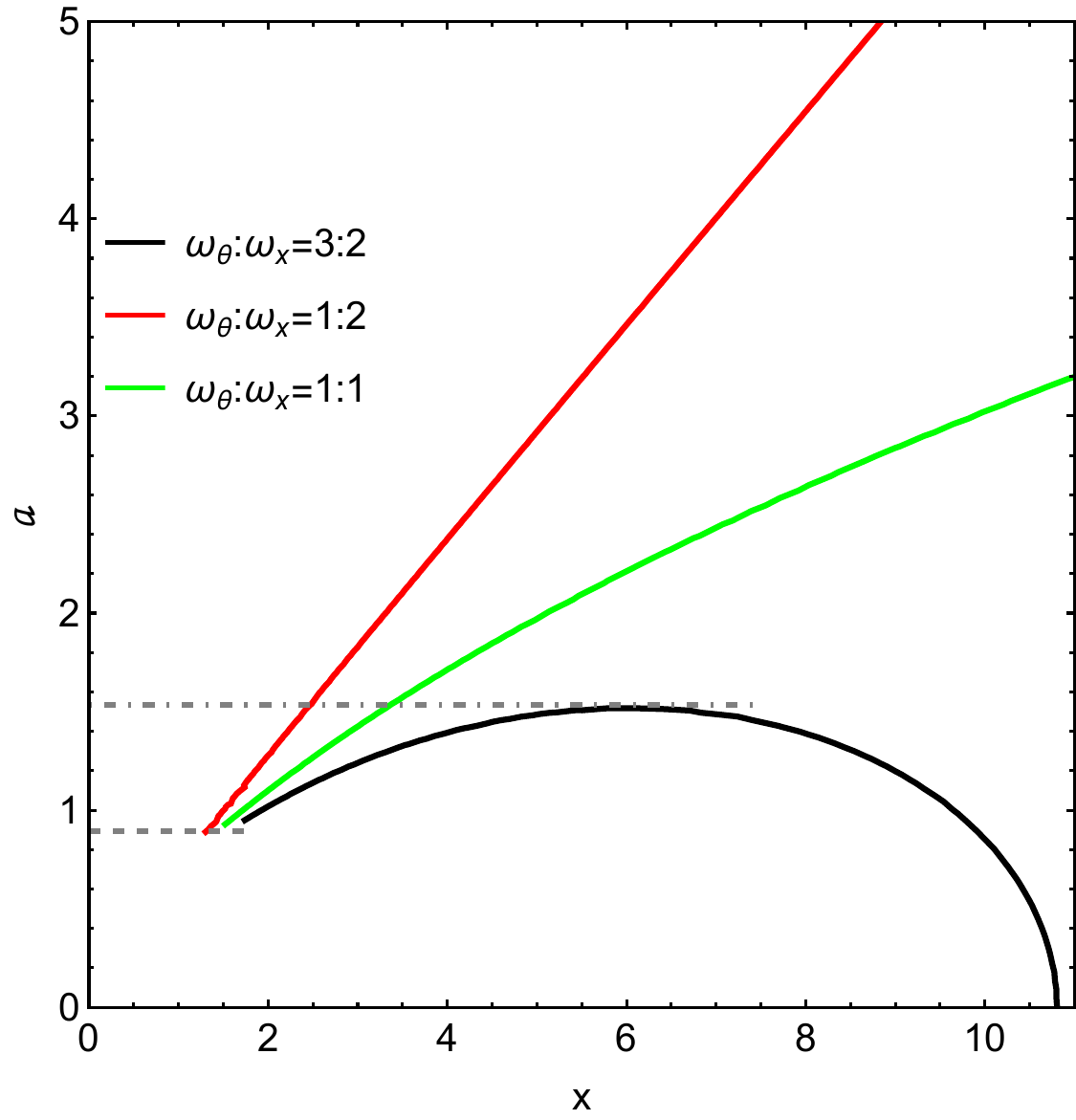}
    \caption{The positions of resonance phenomena for different types of celestial bodies (with different values of $a$) in the parametric resonance model. The dashed line and dotted dashed line correspond to positions on the $y$-axis labeled as $a=2 \sqrt{5} / 5$ and $a \approx 1.534$, respectively.}
    \label{fig:5}
\end{figure}

By selecting specific values of $n$ in the resonance model (e.g., for cases where resonance is more pronounced: $n=1,2,3)$, we plotted the variation of resonance positions $x$ with respect to the parameter $a$ in Figure \ref{fig:5}. From Figure \ref{fig:5}, we can visually observe the positions where resonance occurs for particles around different types of celestial bodies in the BBT spacetime. Specifically, when $n=1,2$ (corresponding to $\omega_\theta: \omega_x=1: 2$ and $\left.\omega_\theta: \omega_x=1: 1\right)$, these two resonance behaviors can only occur in traversable wormholes with larger throats $(a>2 \sqrt{5} / 5)$, and the positions of resonance occurrence move farther away from the center of the radial coordinate as a increases. When $n=3$ (corresponding to $\omega_\theta: \omega_x=3: 2$ ), this resonance mode requires $a \lesssim 1.534$. Additionally, for the case of $0 \leq a \leq 2 \sqrt{5} / 5$, the positions of resonance occurrence move closer to the center of the radial coordinate as $a$ increases. In the case of $2 \sqrt{5} / 5 \leq a \lesssim 1.534$, we observed that resonance phenomena corresponding to the same value of  $a$  can occur at two different positions. The specific reason for this phenomenon is not yet clear, and it may be caused by the unique ring-like structure of the accretion disk around BBT wormholes or different physical processes inside the accretion disk, which requires further exploration.

\begin{figure}[htbp]
  \includegraphics[width=10cm]{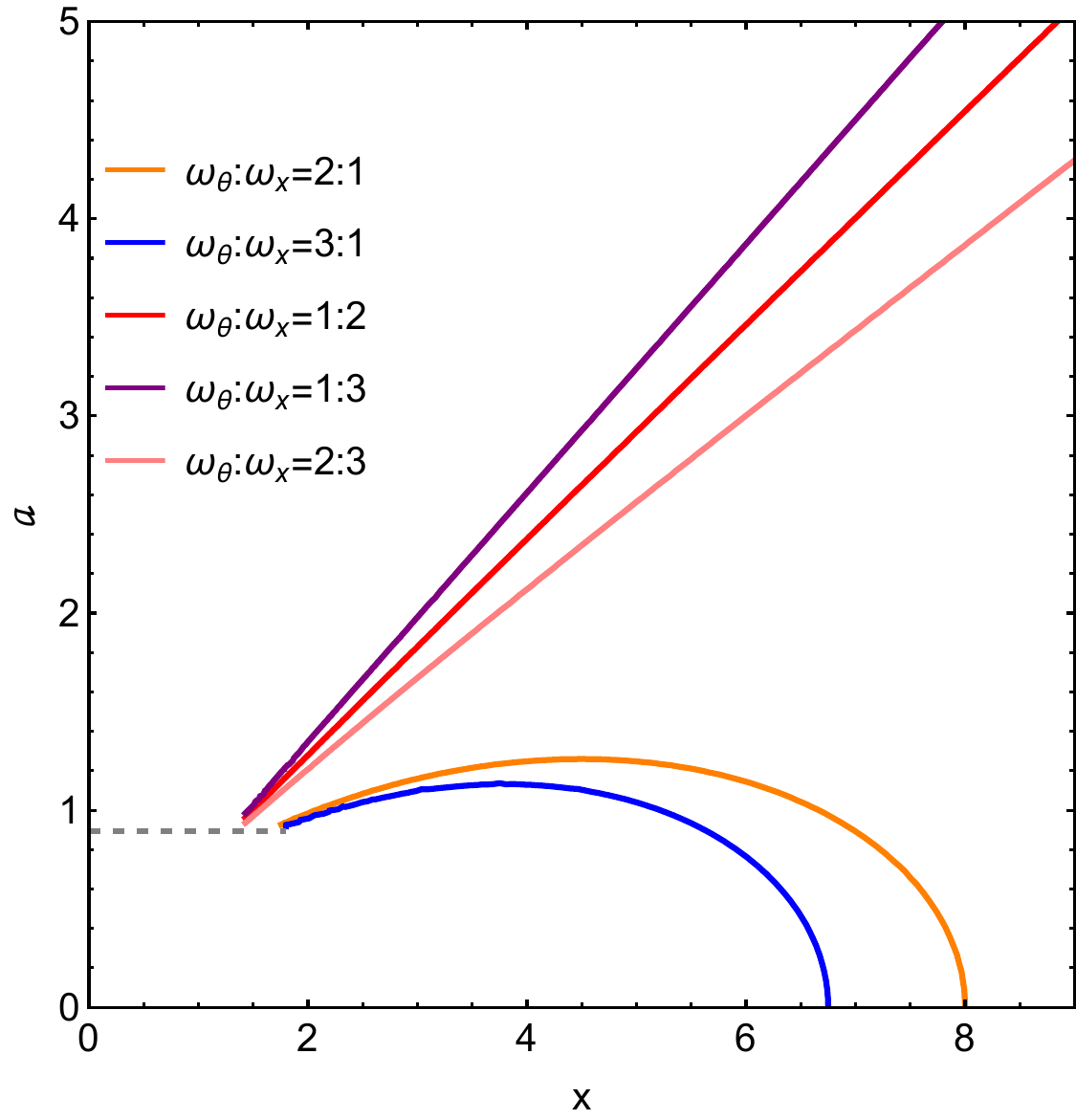}
    \caption{Positions of resonance phenomena for different types of celestial bodies (with different values of $a$) in the forced resonance model. The dashed line corresponds to positions on the $y$-axis labeled as $a=2 \sqrt{5} / 5$.}
    \label{fig:6}
\end{figure}

In practical studies of resonance problems, it is often assumed that factors such as viscous or magnetic stresses in the accretion flow lead to the appearance of non-zero forcing terms \cite{73,77}. Based on this, researchers have established the forced resonance model. In this model, the perturbation equations, which include non-zero forcing terms, can be written as:
\begin{equation}
\delta \ddot{\theta}+\omega_\theta^2 \delta \theta=-\omega_\theta^2 \delta x \delta \theta+F_\theta(\delta \theta),  \label{26}
\end{equation}
where $\delta x=B \cos \left(\omega_{x} t\right), F_\theta$ correspond to the nonlinear terms related to $\delta \theta$. When the relationship between the epicyclic frequencies satisfies the following equation:
\begin{equation}
\frac{\omega_\theta}{\omega_x}=\frac{\nu_\theta}{\nu_x}=\frac{p}{q},   \label{27}
\end{equation}
resonance will be activated. In the equation, $p$ and $q$ are small natural numbers. From Figure \ref{fig:4}, we can see that in the case of $a \leq 2 \sqrt{5} / 5$, we have $\omega_\theta>\omega_{x}$ in the BBT spacetime, which requires $p / q>1$. In this case, prominent resonance phenomena can occur in situations where the frequency ratio is $\omega_\theta: \omega_{x}=p: q=2: 1$ or $3: 1$ (resonance phenomena for $p: q=3: 2$ are the same as parameter resonance for $n=3$, and we won't go into detail here). However, when $a>2 \sqrt{5} / 5$, as we can see from Figure \ref{fig:4}, both $\omega_\theta>\omega_{x}$ and $\omega_\theta \leq \omega_{x}$ can occur, indicating that situations with frequency ratios of $p: q=1: 2, p: q=1: 3$, or $p: q=2: 3$ can induce resonance phenomena. We plotted the variation of resonance positions $x$ with respect to the parameter $a$ in the forced resonance model in Figure \ref{fig:6}.

From Figure \ref{fig:6}, we can visually see that in the BBT spacetime, when $p: q=$
$1: 2, 1: 3$ or $2: 3$ (corresponding to $\omega_\theta: \omega_{x}=1: 2 , 1: 3$ or $ 2: 3$ ), these three cases of resonance phenomena can only occur in traversable wormholes with larger throats $(a>2 \sqrt{5} / 5)$. When $p: q=2: 1$ or $3: 1$, the occurrence of these two resonance modes requires either $a \lesssim 1.271$ or $a \lesssim 1.140$. For the cases of $2 \sqrt{5} / 5 \leq a \lesssim 1.271$ (orange curve) and $2 \sqrt{5} / 5 \leq a \lesssim 1.140$ (blue curve) in the forced resonance model, similar conclusions to those in the parameter resonance model (as shown in Figure \ref{fig:5} for the $2 \sqrt{5} / 5 \leq a \lesssim 1.534$ case) can be drawn. This means that for the same parametric value $a$, the same type of vibration can occur at different positions in the accretion disk.

\section{$\text{Fitting observed data to constrain BBT model and exploring potential mechanisms for producing HFQPOs}$}

It is a well-known fact that in the data of experimentally observed HFQPOs, the high and low frequencies in the double peaks frequently exhibit a fixed ratio of 3:2. Reference \cite{19} studied particles oscillating around a central celestial body in the black-bounce spacetime with resonance models, and discovered that achieving the 3:2 structure observed in microquasars, such as GRO 1655-40, XTE 1550-564, and GRS 1915+105, is not possible. In this section, we explore the frequencies of epicyclic motion of oscillating particles in the BBT geometry and compare our findings with the 3:2 pattern in HFQPOs observed in microquasars. We investigate the various celestial bodies that microquasars could correspond to and assess the potential mechanisms responsible for producing HFQPOs. In addition, we further limit the BBT theoretical model by fitting it with microquasar data.

In order to establish a connection between the theoretical values of the epicyclic motion angular frequencies $\omega$ for particle's local motion and the observed values, we use the redshift factor to transform equations (\ref{20}) and (\ref{21}) as follows:
\begin{equation}
\bar{\omega}=\frac{\omega_{x, \theta}}{-g^{t t} E},  \label{28}
\end{equation}
the expression for $E$ can be found in equation (\ref{9.5}). To ensure that the physical quantities in the theoretical model have the same dimensions as the corresponding observed quantities, we define:
\begin{equation}
\nu_i=\frac{1}{2 \pi} \frac{c^3}{G M} \bar{\omega}_i,~(i=x, \theta)  \label{30}
\end{equation}
where $c$ is the speed of light, $G$ is the gravitational constant, and $M$ is the mass of the celestial body.

\subsection{{Studying on the resonance positions based on the HFQPOs model}}

We consider the observational data of HFQPOs from three sets of microquasars (as listed in Table 1) \cite{78,79}, which are labeled as GRO 1655-40, XTE 1550-564, and GRS 1915+105. The specific data includes the high and low frequencies in the HFOPOs double peaks, the mass of the central celestial body $M / M_{\odot}$, and its spin  $\xi$. Next, we will apply the observational data listed in Table 1 to constrain and analyze the BBT theory.
\begin{table}[ht]
\begin{center}
\begin{tabular}{|c||l|l|l|}
\hline   & GRO 1655-40  & XTE 1550-564 &  GRS 1915+105
\\ \hline $\nu_u[\mathrm{Hz}]$ & 447-453  &273-279 &165-171
\\ \hline $\nu_l[\mathrm{Hz}]$ &295-305   &179-189 &108-118
\\ \hline $M / M_{\odot}$ & 6.03-6.57  &8.5-9.7 &9.6-18.4
\\ \hline $\xi$ &0.65-0.75  &0.29-0.52 &0.98-1
\\ \hline
\end{tabular}
\end{center}
\caption{\label{table-grb-data}
  Observational HFQPOs data for three sets of microquasars.}
\end{table}

\begin{figure}[htbp]
  \centering
    \includegraphics[width=8cm]{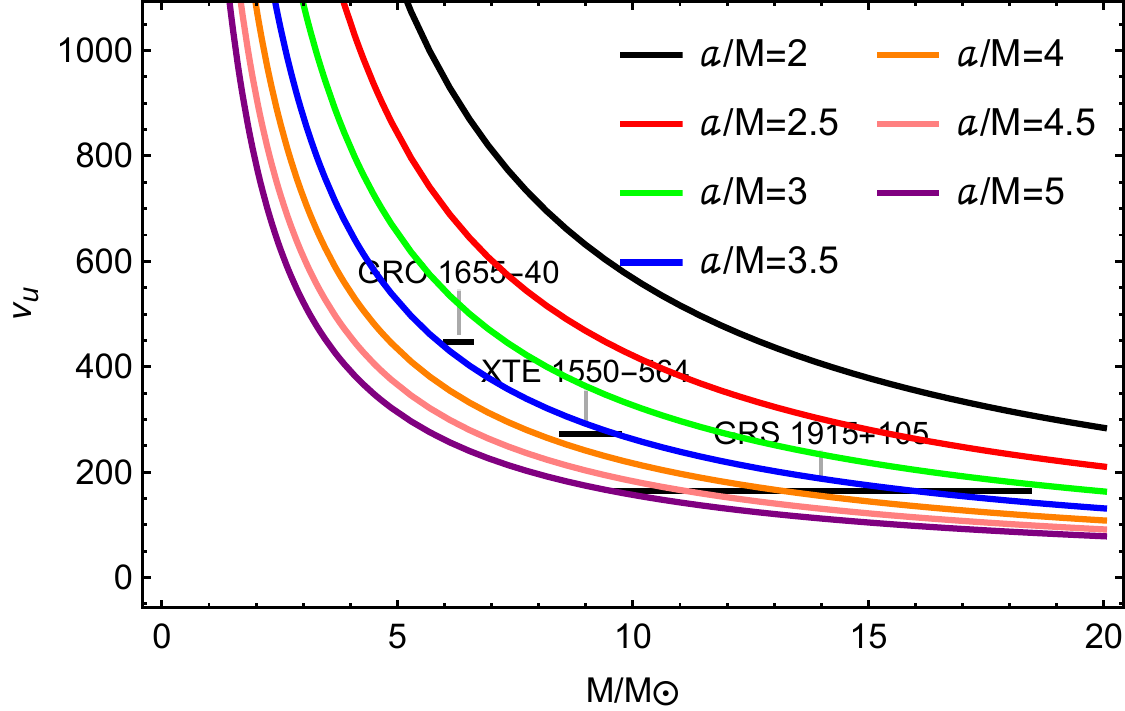} \\
    \includegraphics[width=8cm]{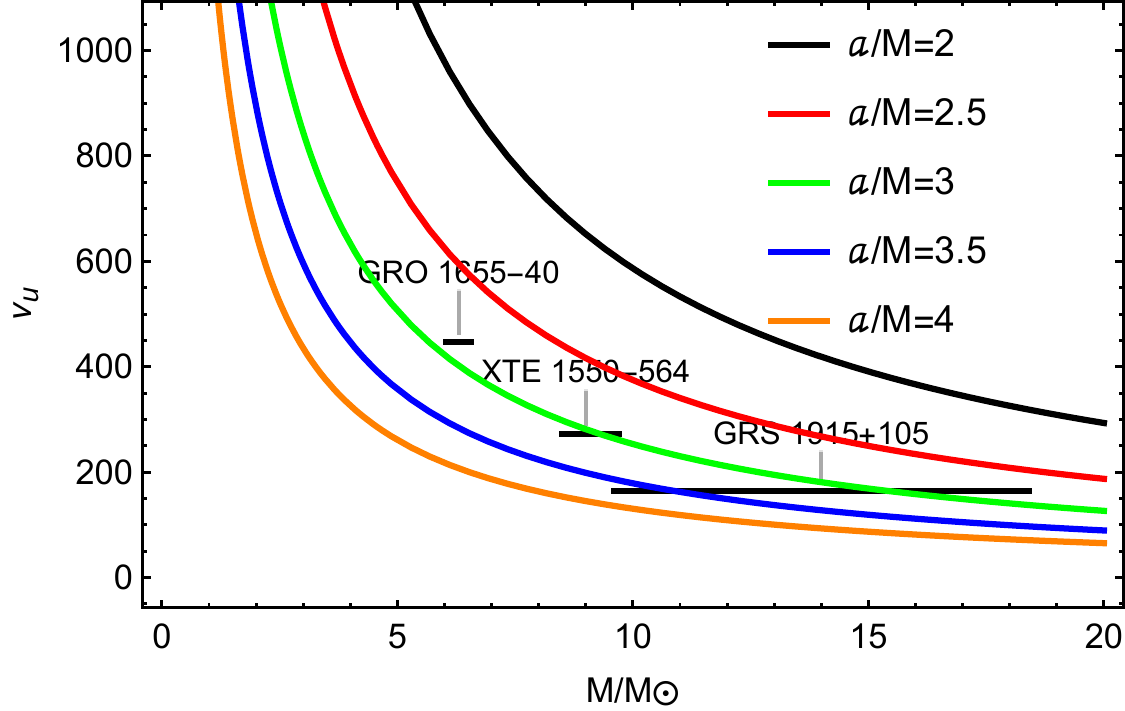}
    \includegraphics[width=8cm]{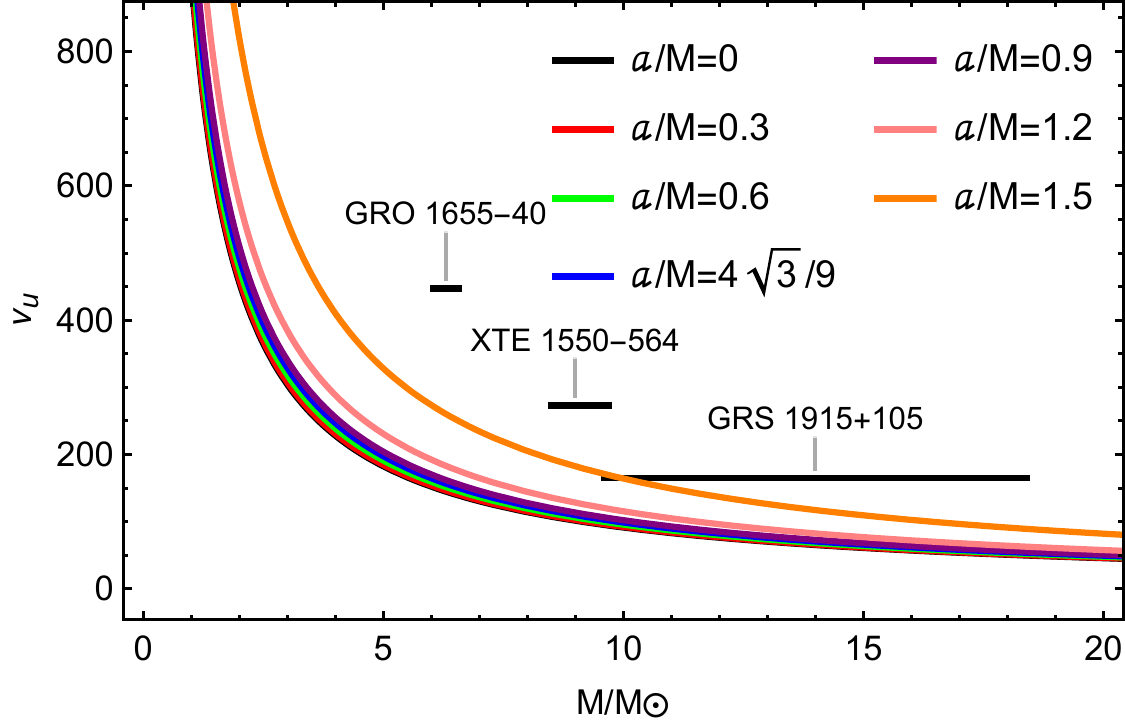}
    \caption{Variation of particle oscillation frequencies relative to the mass of the central celestial body in the BBT spacetime at the resonance point $\nu_{u}: \nu_{l}=3: 2$  with taking different values of $a/M$ in the parametric resonance model. Here, resonance parameters $n=1,2,3$ correspond to the top, bottom left, and bottom right, respectively. The observational data for three sets of microquasars are also displayed in the figure.}
    \label{fig:7}
\end{figure}

Firstly, let's consider the popular parametric resonance model. In Figure \ref{fig:7}, we calculate the resonance frequencies for particles in the BBT spacetime when they oscillate around different types of central celestial bodies. To ensure that the parametric $n$ can achieve the observed result of $\nu_u: \nu_l=3: 2$ for different values of $n$ (e.g., $n=1, 2, 3$ ), we need to consider the possible correspondence between the observed high and low frequencies of the double peaks and the theoretical epicyclic frequencies. In fact, through calculations, it can be found that for a given $n$ value, the ratio of radial to azimuthal frequencies will be determined, and as a result, the resonance positions and the results of applying observational data to constrain the theoretical model will remain unchanged. As an example, in this paper, we consider the following cases for discussion: when $n=1, \nu_u=3 \nu_\theta, \nu_l=\nu_x$; when $n=2$, $\nu_{u}=3 \nu_\theta, \nu_{l}=2 \nu_{x}$; when $n=3, \nu_{u}=\nu_\theta, \nu_{l}=\nu_{x}$.
In addition,  the three-sets observational HFQPOs data form microquasars listed in Table 1 are plotted in Figure \ref{fig:7}, and which are compared with the theoretical values calculated by using the BBT model. We find that under different values of $n$, in order for the theoretical model to pass the experimental observations of microquasars, the constraints on the model parameter $a/M$ with respect to the observational data need to satisfy the results shown in Table 2.

\begin{table}[ht]
\begin{center}
\begin{tabular}{|c||l|l|l|}
\hline   & GRO 1655-40  & XTE 1550-564 &  GRS 1915+105
\\ \hline $n$=1& 3.21-3.47  &3.47-3.83 &3.15-5
\\ \hline $n$=2&2.80-2.92   &2.94-3.13 &2.76-3.69
\\ \hline $n$=3& ---  &--- &1.49-1.53
\\ \hline
\end{tabular}
\end{center}
\caption{\label{table-grb-data}
 Constraints on the BBT model parameter $a/M$ from the observational data of three sets of microquasars under different resonance parameter values $n$ in the parameter resonance model.}
\end{table}

From Figure \ref{fig:7}, we can see that the oscillation frequencies of particles located on stable circular orbits in the BBT spacetime can closely match the observational data of the three microquasars when the resonance parameter is set to $n=1$ or $n=2$ (e.g., when $n=1, a/M=3.5$, and when $n=2, a/M=3$). This indicates that the observed resonance phenomena can also be generated by particles oscillating around a central celestial body as a wormhole $(a/M>4 \sqrt{3} / 9)$ in the BBT spacetime. However, when $n=3$, the BBT model deviates significantly from the observational data. Table 2 presents the constraint results of fitting the observational data under  assumptions for different frequency ratio to the model parameter $a/M$. Obviously,  for the cases of $n=1$ and $n=2$, the fitting results suggest that the central celestial body corresponds to a wormhole. Furthermore, from Table 2, it can be found that the constraint value of the model parameter $a/M$ for $n=1$ is greater than the fitting value of $a/M$ for $n=2$. Combining Table 2 and Figure \ref{fig:5}, we can conclude that for both $n=1$ and $n=2$, the resonance occurs near the throat of the wormhole, making QPOs phenomena a tool for probing strong gravity effects.

\subsection{{Data fitting based on the forced resonance model and results}}

\begin{figure}[htbp]
  \includegraphics[width=5.5cm]{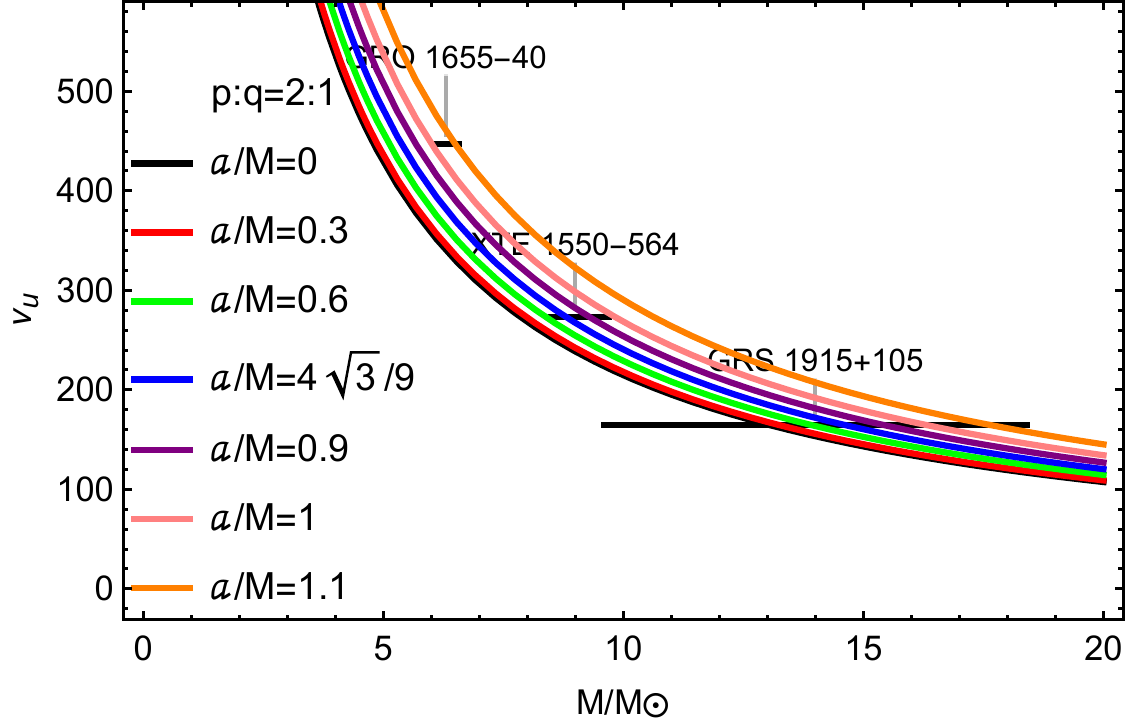}
  \includegraphics[width=5.5cm]{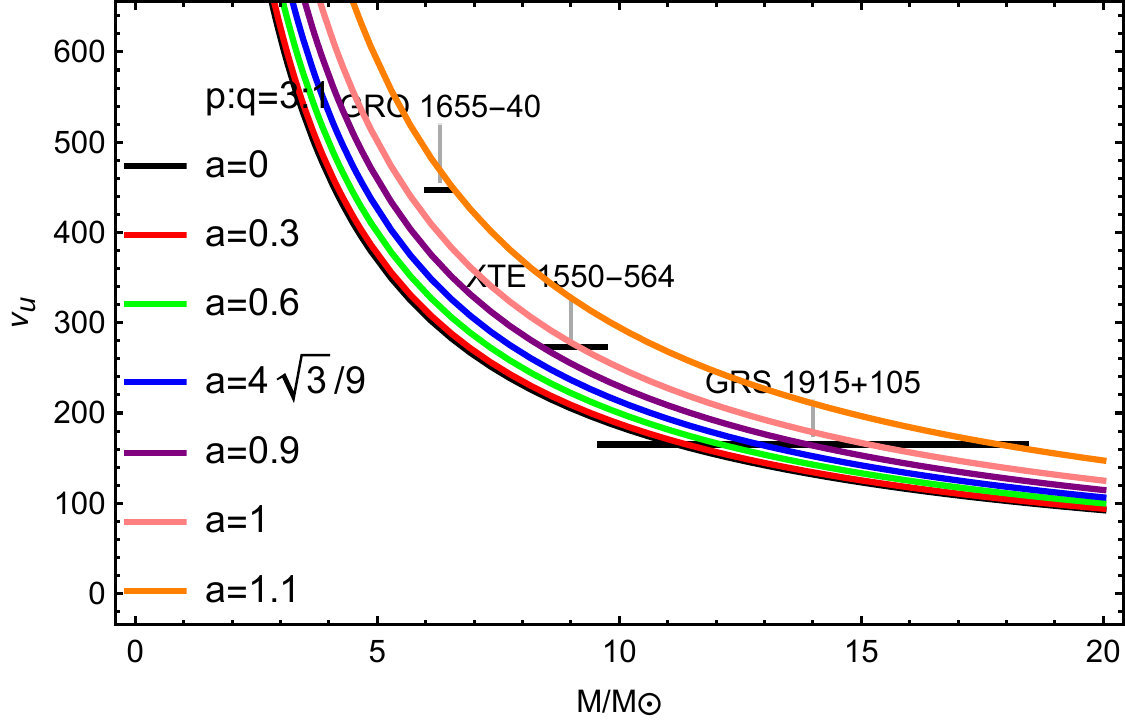}
  \includegraphics[width=5.5cm]{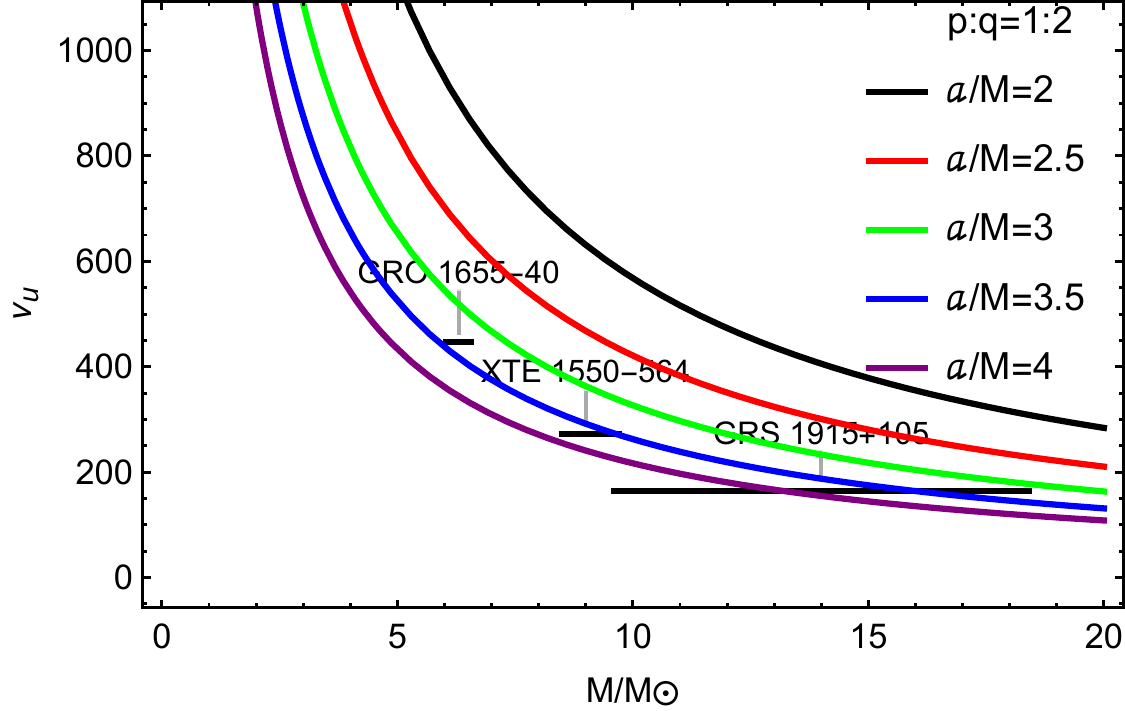}
  \includegraphics[width=5.5cm]{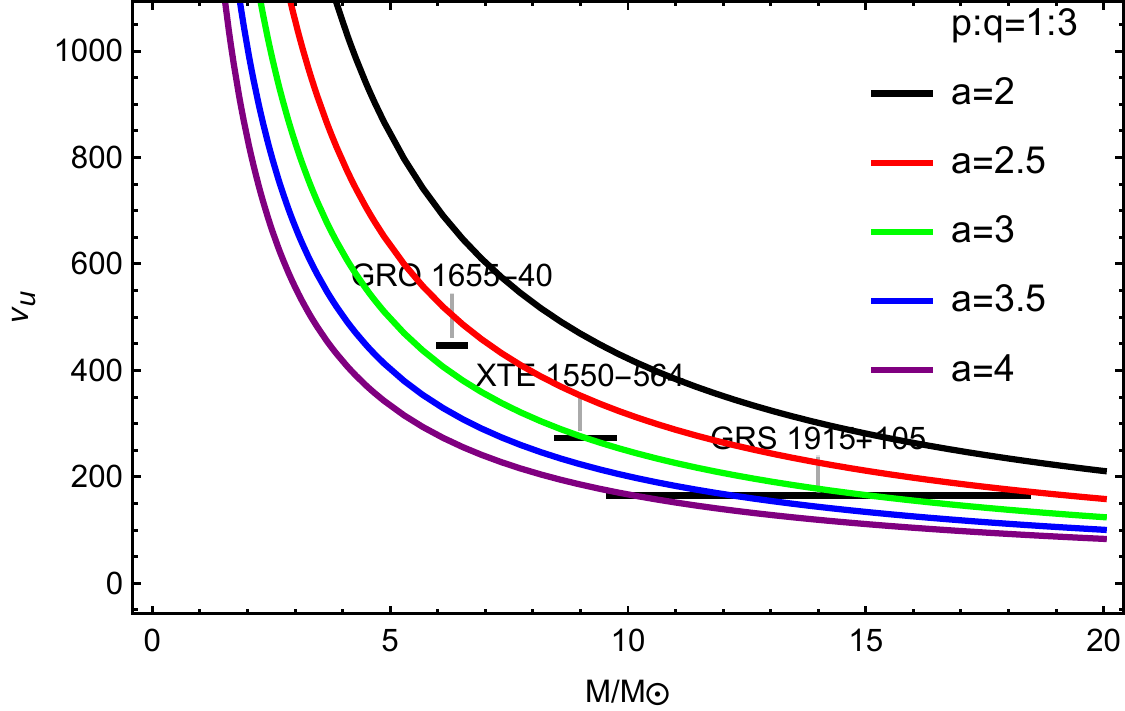}
  \includegraphics[width=5.5cm]{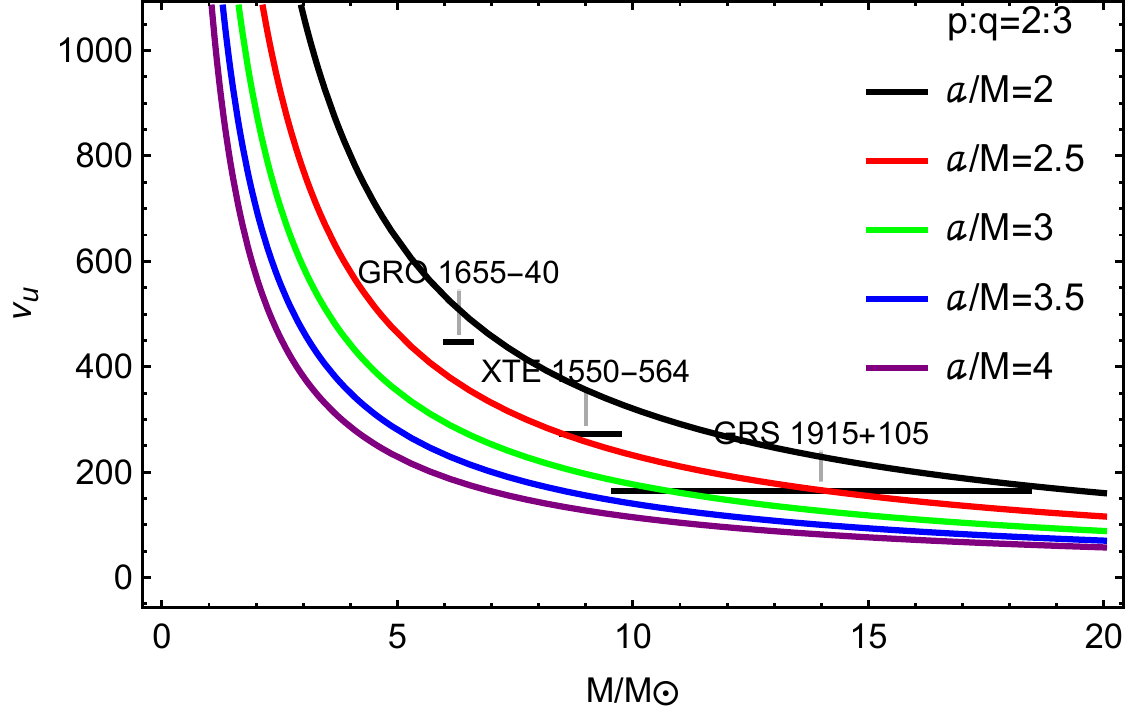}
    \caption{Variations of particle oscillation frequencies relative to the mass of the central body in BBT spacetime at $\nu_{u}: \nu_{l}=3: 2$ within the forced resonance model, where the different values of $a/M$ are taken. We also compare these theoretical results with observational data.}
    \label{fig:8}
\end{figure}

For models focusing on the relationship between the radial and latitudinal oscillation frequencies, there are typically two types: the parametric resonance model and the forced resonance model. In this section, to analyze other potential mechanisms for generating HFQPOs in the BBT spacetime, we apply observational data to constrain and test the theoretical model based on the forced resonance hypothesis. Similarly, in order to ensure the double peak structure of $v_{u}: \nu_{l}=3: 2$ under different $p:q$ ratios in the forced resonance model, we consider the following theoretical expressions for $\nu_{u}$ and $\nu_{l}$. For example, when $p:q=2: 1, \nu_{u}=\nu_\theta+\nu_{x}$, $\nu_{l}=\nu_\theta$; when $p: q=3: 1, \nu_{u}=\nu_\theta, \nu_{l}=\nu_\theta-\nu_{x}$; when $p:q=1: 2, \nu_{u}=\nu_\theta+\nu_{x}$, $\nu_{l}=\nu_{x}$; when $p:q=1: 3, \nu_{u}=\nu_{x}, \nu_{l}=\nu_{x}-\nu_\theta$; when $p:q=3: 2, \nu_{u}=\nu_{x}, \nu_{l}=$ $\nu_\theta$. In Figure \ref{fig:8}, we compare the theoretically calculated frequency values based on the forced resonance in the BBT spacetime with the observational data from microquasars. We also use astronomical experimental data to constrain the model parameter $a/M$ (results are shown in Table 3).

\begin{table}[ht]
\begin{center}
\begin{tabular}{|c||l|l|l|}
\hline   & GRO 1655-40  & XTE 1550-564 &  GRS 1915+105
\\ \hline $p:q$=2: 1& 1-1.12  &0.62-0.99 &0-1.143
\\ \hline $p:q$=3: 1&1.05-1.1   &0.9-1.05 &0-1.11
\\ \hline $p:q$=1: 2&3.23-3.45 &3.48-3.83 &3.16-4.98
\\ \hline $p:q$=1: 3&2.65-2.85&2.85-3.16&2.58-4.16
\\ \hline $p:q$=2: 3&2.12-2.26&2.28-2.5&2.07-3.23
\\ \hline
\end{tabular}
\end{center}
\caption{\label{table-grb-data}
 Constraints on the BBT model parameter $a/M$ from the three-sets observational microquasars data under the forced resonance models.}
\end{table}

Based on the constraints provided by fitting the microquasars data (Table 3), we find that the resonant phenomena excited in the BBT theory can be explained through the forced resonance model. Specifically, we observe that: for the frequency ratio $p:q=2: 1,$ black hole in BBT spacetime $(a/M \leq 4 \sqrt{3} / 9)$ can be tested against the observational data from XTE 1550-564 and GRS 1915+105. For the case of $p:q=3: 1$, the quasi-periodic oscillations of particles around black holes in BBT spacetime align with the observations of microquasar GRS 1915+105. Moreover, in the BBT wormhole spacetime $(a/M>4 \sqrt{3} / 9)$, for cases of taking some specific values of $p:q$ listed in Table 3, the BBT model can meet the requirements tested by the observations of the three types of microquasars. This suggests that the observed oscillatory behavior in these three microquasar classes can be explained by particle oscillations occurring in the BBT wormhole spacetime.

\section{$\text{Conclusion}$}

Regular black holes were proposed as a solution to the spacetime singularity problem in gravitational physics. The BBT spacetime metric, as proposed by Lobo et al., has the capability to describe various objects such as Schwarzschild BH, regular BH, extremal BH, and traversable wormhole, depending on the varying values of the model parameter $a$.  Following the method shown in Ref.\cite{BB-action-prd,BBT-action-prd}, it is found that the BBT solution can be obtained by Einstein’s theory of general relativity sourced by a combination of a minimally coupled self-interacting phantom scalar field with a nonzero potential and a nonlinear electromagnetic field.

 In the BBT spacetime studied in this paper, we explored the regions of stable circular orbits and investigated the locations of the ISCOs for various celestial bodies. Research indicates that for both regular black holes and extremal black holes, only a single ISCO exists. In contrast, traversable wormholes can exhibit either one or two ISCOs, depending on the size of the throat. Furthermore, as QPOs are potent tools for testing gravitational theories, our research concentration was placed on particles oscillating on stable circular orbits around central bodies. We investigated the properties of the angular frequencies of their radial and latitudinal epicyclics. It is shown that particles surrounding various types of celestial bodies display unique frequency oscillation characteristics. When the BBT spacetime describes black holes and wormholes with single or double photon spheres $(0 \leq a \leq 2 \sqrt{5} / 5)$, particles in the region of $x \geq x_{\text {ISCO }}$ demonstrate a higher radial epicycle frequency than their latitudinal epicycle frequency. The epicycle frequency characteristics in these scenarios resemble those of Schwarzschild black hole, wherein the latitudinal frequencies decrease monotonically with increasing radial coordinates $x$ and possess a single-peaked structure. In contrast, for wormholes with a single photon sphere $(a>2 \sqrt{5} / 5)$, there is a result where radial epicycle frequencies are greater than latitudinal epicycle frequencies (in contrast to the results in black hole spacetime), and the epicycle frequency differ significantly from those in Schwarzschild BH, this means that lower-order resonance parameters can be excited, resulting in stronger observational signals.

The research on the phenomenon of HFQPOs generated by particles around wormholes using microquasar data is still limited. This paper conducts a theoretical study by fitting observational data within the framework of spacetime metrics capable of describing both black holes and wormholes simultaneously. Using two resonance models, we offer numerical calculations of resonance occurrence positions in the BBT spacetime for various celestial bodies (differing in $a$-values) in relation to their corresponding frequencies. Furthermore, we investigate the possibility of utilizing the oscillation data from three microquasars to assess the feasibility of testing the BBT model. The research reveals that the resonance positions move away from the central origin as the value of $a$ increases when the resonance parametric $n=1$ or 2, for the case of $a>2 \sqrt{5} / 5$. Conversely, in the case of $0 \leq a \leq 2 \sqrt{5} / 5$, the resonance positions shift closer to the central origin as the value of parameter $a$ increases. Moreover, the research suggests that when parametric resonance is triggered (e.g., $n=1$ or 2 ), the observable aligns closely with the traversable wormhole model in the BBT spacetime ( $a>4 \sqrt{3} / 9)$. And in the forced resonance models, black hole or wormhole models can be tested through observations at different frequency ratios in the radial and latitudinal directions.

Finally, we used observational data to constrain the regularization parameter $a/M$ in the BBT spacetime (results detailed in Tables 2 and 3) and analyzed the possible mechanisms for the generation of HFQPOs. The study found that, unlike the black bounce spacetime, which cannot be tested by microquasar observation data under the resonance model \cite{19}, in the BBT spacetime, the oscillatory behavior of three types of microquasars can also be explained by the particle oscillation phenomenon that occurs in the BBT spacetime under the parameter resonance and forced resonance models. This means that the BBT model improves the poor fit between the black-bounce spacetime and microquasar observational data, while also providing a basis for exploring the existence of wormholes.

\textbf{\ Acknowledgments }
 The research work is supported by   the National Natural Science Foundation of China (12175095,12075109 and 11865012), and supported by  LiaoNing Revitalization Talents Program (XLYC2007047).

\end{document}